\documentclass[11pt,twoside]{article}
\usepackage[makeroom]{cancel}
\usepackage{latexsym}
\usepackage{longtable}
\usepackage{epsfig}
\usepackage{graphicx,bbm,psfrag}
\usepackage{amssymb,amsmath,amsthm,booktabs,mathtools}
\usepackage{bm}
\usepackage{color}
\usepackage{xcolor}
\usepackage{dutchcal}
\usepackage{hyperref}
\hypersetup{
    colorlinks,
    citecolor=red,
    linkcolor=blue,
}

\newtheorem{theorem}{Theorem}[section]

\newcommand{\bc}{\begin{center}}
\newcommand{\ec}{\end{center}}
\def\ba#1{\begin{array}{#1}\displaystyle}
\newcommand{\ea}{\end{array}}

\newcommand{\beq}{\begin{equation}}
\newcommand{\eeq}{\end{equation}}
\newcommand{\beqa}{\begin{eqnarray}}
\newcommand{\eeqa}{\end{eqnarray}}

\newcommand{\n}{\nonumber\\}
\newcommand{\bi}{\begin{itemize}}
\newcommand{\ei}{\end{itemize}}

\renewcommand{\v}[1]{\boldsymbol{#1}}

\def\t#1{\tilde{#1}}

\def\b#1{\bar{#1}}
\def\frc#1#2{\frac{#1}{#2}}

\newcommand{\p}{\partial}

\newcommand{\bra}{\langle}
\newcommand{\ket}{\rangle}
\newcommand{\Z}{{\mathbb{Z}}}

\newcommand{\R}{{\mathbb{R}}}
\newcommand{\C}{{\mathbb{C}}}
\newcommand{\Tr}{{\rm Tr}}

\newcommand{\Or}{{\cal o}}

\newcommand{\ep}{\epsilon}

\newcommand{\ri}{{\rm i}}

\newcommand{\dd}{{\rm d}}

\begin{document}

\begin{titlepage}

\begin{center}
{\Large {\bf Free energy fluxes  \\[0.1cm] and the Kubo-Martin-Schwinger relation }}

\vspace{1cm}

{\large Benjamin Doyon and Joseph Durnin}
\vspace{0.2cm}

{\small\em
Department of Mathematics, King's College London, Strand, London WC2R 2LS, UK}

\end{center}
A general, multi-component Eulerian fluid theory is a set of nonlinear, hyperbolic partial differential equations. However, if the fluid is to be the large-scale description of a short-range many-body system, further constraints arise on the structure of these equations. Here we derive one such constraint, pertaining to the free energy fluxes. The free energy fluxes  generate expectation values of currents, akin to the specific free energy generating conserved densities. They fix the equations of state and the Euler-scale hydrodynamics, and are simply related to the entropy currents. Using the Kubo-Martin-Schwinger relations associated to many conserved quantities, in quantum and classical systems, we show that the associated free energy fluxes are perpendicular to the vector of inverse temperatures characterising the state. This implies that all entropy currents can be expressed as averages of local observables. In few-component fluids, it implies that the averages of currents follow from the specific free energy alone, without the use of Galilean or relativistic invariance. In integrable models, in implies that the thermodynamic Bethe ansatz must satisfy a unitarity condition. The relation also guarantees physical consistency of the Euler hydrodynamics in spatially-inhomogeneous, macroscopic external fields, as it implies conservation of entropy, and the local-density approximated Gibbs form of stationarity states. The main result on free energy fluxes is based on general properties such as clustering, and we show that it is mathematically rigorous in quantum spin chains.
\vfill

\hfill \today

\end{titlepage}

\tableofcontents

\section{Introduction}

It is a striking property of a large class of thermodynamically large systems, that the state reached after relaxation allows the expectation values of all observables to be determined by the values of a few conserved quantities. The interpretation is simple: these states maximise entropy with respect to the few extensive constraints imposed by the dynamics. For instance, if the only such constraint is energy conservation, stationary states of the system are well described by the canonical Gibbs ensemble, with density function $\rho=e^{-\beta H}$, where $H$ is the Hamiltonian and $\beta$ the associated Lagrange parameter (or potential) identified with the inverse temperature. In general, within appropriate ranges, by thermodynamic convexity the relationship between the potentials and the charge densities is a bijection. Thus all physical quantities are equivalently functions of the potentials, or the densities; for instance, the expectation values of currents, as functions of densities, are model-dependent functions known as the equations of state. The reduction of the number of degrees of freedom, and the equations of states, are the crucial ingredients in the associated Euler hydrodynamic equations, which describe the large-scale dynamics and follows from local thermodynamic relaxation. The more difficult question of whether such states are actually reached dynamically is the subject of ergodic theory in classical systems \cite{khinchin1949mathematical}, and thermalisation and the eigenstate thermalisation hypothesis in quantum systems \cite{DallesioETHreview2016,eisertreview}.

The usefulness of descriptions based on few conserved quantities is apparent from the fact that it is possible to accurately reproduce the large-scale dynamics of complicated systems, such as a jet engine, with the knowledge of few observables such as the pressure, temperature and density. However, there is an important class of systems where the maximal-entropy principle does not generally reduce the degrees of freedom to such a small set of observables: integrable models, both quantum and classical. These admit a space of extensive conserved quantities that grows with the size of the system. The thermodynamics of integrable models is correspondingly richer. Maximising entropy with respect to these conserved charges lead to Generalised Gibbs Ensembles (GGEs) \cite{1742-5468-2016-6-064002,vidmar2016generalized,IlievskietalQuasilocal,GogolinEquilibration2016}. A powerful framework for their thermodynamics is the thermodynamic Bethe ansatz \cite{Yang-Yang-1969,ZAMOLODCHIKOV1990695,TakahashiTBAbook}, which can be shown to lead to a qualitatively different Euler scale hydrodynamics \cite{PhysRevLett.117.207201,PhysRevX.6.041065,PhysRevB.97.045407,DoyonLecture2020} (referred to as generalised hydrodynamics). GGEs and their hydrodynamics are relevant to subjects as widely separated as cold atomic gases \cite{Langen207,Schemmerghd} and classical soliton gases \cite{CDE16}, and to the dynamics of systems which are close to being integrable \cite{langen2016prethermalization} (see the recent works \cite{mallayya_prethermalization_2018,vas19,DurninTherma2020,lopez2020hydrodynamics,bastianello2020noise,bouchoule2020effect,moller2020extension,bastianello2020thermalisation}).

A crucial observation, that has been made apparent in the studies of integrable systems but that is more general, is that, regardless of the number of conserved quantities (finite or extensive) and of the type of system, the same maximal-entropy principle seems to hold, with a high degree of universality. That is, at long times, the states maximise entropy with respect to the complete set of all extensive conserved quantities $\{Q_i = \int \dd^d x\,q_i(\v x)\}$. The formal density function therefore takes the form
\beq\label{rho}
	\rho\propto e^{-\sum_i\beta^iQ_i},
\eeq
where in the integrable case the series in the exponential can contain an infinity of terms, under a suitable notion of convergence and completeness (see e.g. discussions in \cite{GogolinEquilibration2016,IlievskiInteracPart,Doyon2017,pozsgay2017generalized}). We will refer to such states as maximal entropy states, the constraints given by the conserved quantities being implied.

The form \eqref{rho} defines, for both chaotic and integrable systems, the universal ensemble description underpinning the thermodynamics and the Euler hydrodynamics of many-body models. Indeed, the thermodynamics is the theory for average conserved densities $\bra q_i\ket$ in such states, while the Euler hydrodynamics is the dynamics induced by the local conservation laws:
\beq\label{hydro}
	\p_t\bra q_i(\v{x},t)\ket + \nabla\cdot  \bra\v{j}_i(\v{x},t)\ket=0,
\eeq
supplemented by the equations of state $\bra\v{j}_i(\v{x},t)\ket = \bra \v{j}_i\ket(\{\bra q_j(\v{x},t)\ket\})$ for the family of maximal entropy states \eqref{rho}. In many-body models with short-range interactions, the form \eqref{rho} imposes constraints on the large-scale degrees of freedom and their dynamics, beyond the otherwise arbitrary hyperbolic form \eqref{hydro}. One particular well known example of such a constraint is the fluctuation-dissipation theorem. In the context of non-equilibrium physics, it is desirable to establish which properties hold at the Euler scale due to this structure.

It turns out that the Gibbs form \eqref{rho} of maximal entropy states implies the existence of a free energy flux $\v{g}$ which generates the currents $\bra\v{j}_i\ket = \p\v{g}/\p\beta^i$, see \cite{PhysRevX.6.041065,DoyonLecture2020}. This parallels the thermodynamic relation $\bra q_i\ket = \p f/\p\beta^i$ for the (appropriately normalised) specific free energy $f$. The free energy flux also enters the universal expression for the entropy current \cite{DoyonLecture2020}. In fact, a free energy flux $\v{g}_k$ exists for every conserved quantity $Q_k$ in the model, as each of these quantities (assumed here to be in involution) generates a separate flow to which a set of currents $\v{j}_{ki}$ can be assigned. The generalised currents are physically relevant, for example, in the hydrodynamics of systems where external fields such as external forces or varying interacting strengths are present \cite{DY,BasGeneralised2019,bastianello2020thermalisation}.

In this paper we show, under extremely general hypotheses, that the free energy fluxes must satisfy the projection equation
\beq\label{gintro}
	\sum_k\beta^k \v{g}_k = \v G,
\eeq
where $\v G$ is a constant vector, and $\v G = \v 0$ if the model admits a parity-symmetric conserved quantity.

This relation has a number of consequences. In particular it implies that free energy fluxes and entropy currents can be expressed in terms of averages of local observables. Strikingly, when a single energy-like conserved quantity is considered (in addition, possibly, to the momentum and ultra-local conserved quantities such as the number of particles), the relation \eqref{gintro} gives the {\em exact values of average currents} solely from the knowledge of specific free energy $f$, without the need for Galilean, relativistic or other space-time boost invariances. Fixing the equations of state is in general a hard problem, but with a single energy-like conserved quantity, the knowledge of the free energy suffices. Eq.~\eqref{gintro} further implies a unitarity constraint on the structure of the thermodynamic Bethe ansatz in integrable models. Perhaps most importantly, it guarantees {\em physical consistency of the emergent Euler-scale hydrodynamic equations with spatially-varying external fields}, including the Gibbs form of stationarity states and entropy conservation.

The result \eqref{gintro} is a consequence of the Kubo-Martin-Schwinger (KMS) relation, which characterises (generalised) Gibbs states both in the quantum \cite{IsraelConvexity,BratelliRobinson12} and classical \cite{aizenman1977equivalence} settings\footnote{As typically formulated the KMS relation refers to thermal Gibbs ensembles, but a generalised form of the KMS relation, which takes into account all the conserved quantities and applies to the GGEs, is easily formulated.}. In particular, we show how in quantum spin chains, all hypotheses are satisfied, thus in this context the result is mathematically rigorous.

The paper is organised as follows. In section \ref{sectThermo}, we recall the theory of thermodynamic states in extended systems with short-range interactions, in arbitrary dimension. The theory is most rigorously expressed within the $C^*$ algebra formulation, but here we keep the language simpler for readability. We explain our hypotheses. In section \ref{sectMain} we express our main result \eqref{gintro}, and some of its immediate consequences. Thereafter, we refer to the main result as an Euler-scale KMS relation (EKMS). In section \ref{sectAppl} we describe applications of the EKMS relation to systems with few conserved quantities, to integrable systems, and to the Euler-scale hydrodynamic equations of arbitrary systems within external space-varying fields. In section \ref{sectmoment} we show a general relation, valid in thermodynamic states under our hypotheses, which relates the index-symmetric part of the generalised currents to commutators, or Poisson brackets, of conserved densities. Finally, in section \ref{sectproof} we show the EKMS relation. Appendices provide supporting calculations, in particular we show full mathematical rigour of our derivation in the context of quantum spin chains in appendix \ref{app_rigour}.

\section{Extended systems in $d$ dimensions and thermodynamic states}\label{sectThermo}

In this section we describe the general context in which the main results apply. We consider a many-body, extended system in $d$ dimensions of space, with short-range interactions and in infinite volume. The system may be quantum or classical; we will express the conditions in the quantum case, and review how the classical limit is taken. The type of system is arbitrary: it can be a lattice model, a gas of particles, or a field theory, and it can be integrable or not, and possess nontrivial interaction or not. The central objects are the extensive conserved quantities, for instance the total number of particles, the total energy, the total momentum, or higher conserved quantities if the model is integrable.

We make certain explicit hypotheses about the properties of the system. The main results of section \ref{sectMain}, whose derivation is in sections \ref{sectmoment} and \ref{sectproof}, are obtained under these hypotheses.

The emphasis in the main text is on the universality of the results, and how they are based on general properties which are expected to hold in large classes of statistical systems. Hence in the main text we do not attempt full mathematical rigour and a precise framework. However, we comment below on known rigorous results, and we believe that within the framework of the $C^*$ algebra description of quantum lattice models with finite local space \cite{BratelliRobinson12}, the hypotheses made can be verified rigorously. For completeness, in appendix \ref{app_rigour} we show that this is the case in quantum spin chains, where our main results are in fact rigorous, giving the particular example of the Heisenberg spin-$1/2$ chain.

\subsection{Conserved quantities and maximal entropy states}\label{ssectMES}

Let the system possess a set of extensive conserved quantities (or charges) $Q_i$ in involution, $[Q_i,Q_j]=0$ for all $i,j$. Here $i$ lies in some index set; it may be taken to be infinite in integrable models\footnote{It is not necessary to think of $i$ as in index -- it can be thought of as an abstract index notation, as introduced by Penrose.}, although we will not discuss any convergence issue that may arise in this case. In relaxation processes towards thermodynamic states, only a certain category of conserved quantities are physically relevant (excluding for instance, in quantum systems, projections onto eigenstates of the Hamiltonian): those which are extensive. For our purpose, we assume that each charge has an associated density $q_i(\v{x})$ satisfying 
\beq\label{Qi}
	Q_i = \int \dd^d x\,q_i(\v{x}).
\eeq
The charge density $q_i(\v{x})$ is a local observable at the point $\v{x}\in\R^d$. In the discrete case we take $\int \dd^d x\rightarrow \sum_{\v{x}\in\Z^d}$. As a notion of locality, we will assume for simplicity the observable to be supported on a finite region; but the arguments can be extended to include quasi-local observables, whose projections on regions $R\subset \R^d$ away from $\v{x}$ have norms that decay exponentially with the distance of $R$ to $\v{x}$.

Extensive conserved charges can be given a mathematically accurate definition and completed to a Hilbert space \cite{IlievskietalQuasilocal,Doyon2017}, and are rigorously shown to occur in the Boltzmann-Gibbs principle and the linearised Euler equations in quantum spin chains in \cite{DoyonProjection}. Here we do not use this precise formulation, and it is not necessary for the set of charges $Q_i$ that we consider to be complete (to form a basis for the Hilbert space of extensive conserved charges).

There always exists a Hamiltonian which generates time translations, and there may exist a total momentum vector; that is,
\beq\label{gennot}
	H=Q_2\quad\mbox{(Hamiltonian),}\qquad P_\alpha=Q_{1_\alpha},\ \alpha=1,2,\ldots,d \quad\mbox{(total momentum vector).}
\eeq
In this notation, the numerals $0,1,2,\ldots$ organise the conserved charges according to the local extent of their densities, and the index $\alpha$ is the spatial direction\footnote{Throughout, we will use Latin indices enumerating the charges, and Greek spatial indices.}.
In lattice models, due to the discreteness of space there is no microscopic, continuous translation invariance, and hence no microscopically defined momentum operator. We will assume the presence of a conserved momentum mainly for the interpretation of some of the results.

Because the charges are assumed to be in involution, we assume that each conserved quantity $Q_k$ generates a current $\v{j}_{ki}(\v{x})$ by its action on each conserved density $q_i(\v{x})$. These satisfy the continuity equations
\beq\label{evol}
	\frc{\ri}{\hbar}[Q_k, q_i(\v{x})] + \nabla\cdot \v{j}_{ki}(\v{x}) = 0. 
\eeq
Here we use the quantum notation, but a similar equation holds in classical systems; see below. To our knowledge, the ``generalised" currents $\v{j}_{ki}(\v{x})$ were first discussed in the context of integrable models in \cite{DY}, though the concept clearly does not rely on integrability. Again, the same definition holds on discrete space with the discrete derivative $\p  \Or(\v{x})/\p{x^\alpha} \rightarrow \Or(\v{x})- \Or(\v{x}-\v{e}_\alpha)$, where $\v{e}_\alpha$ is a unit vector along the $\alpha$'th direction, $\Or$ is any observable, and $\Or(\v{x})$ is the translation of $\Or$ by $\v{x}\in\Z^d$. We note that in quantum spin chains, the existence of local currents associated to local conserved densities is proven rigorously, as elements of the Gelfand-Naimark-Segal Hilbert space, in \cite{DoyonProjection}. Below, we will denote the physical currents, generated by the Hamiltonian, as
\beq
	\v{j}_{i} = \v{j}_{2i}\qquad\mbox{(physical currents).}
\eeq
In this notation, the spatial components of the stress tensor are
\beq\label{stress}
	{\cal T}^\gamma_\alpha = j^\gamma_{1_\alpha}.
\eeq
As the momentum operator $P_\alpha$ generates space translations, $\p\Or/\p {x^\alpha} = -\ri\hbar^{-1}[P_\alpha,\Or]$, then \eqref{evol} shows that we can choose, for the currents with respect to this generator,
\beq\label{j1i}
	\v{j}_{1_\alpha i}=\v{e}_{\alpha} q_i.
\eeq
 
The states of interest are the maximal entropy states (MES). Physically, these are states which emerge after relaxation in infinite-volume systems. There are strong indications that extensive conserved charges are sufficient in order to characterise MES's \cite{GogolinEquilibration2016,IlievskiInteracPart,Doyon2017,pozsgay2017generalized}.  They are expected to take the Gibbs form \eqref{rho}; more precisely, the infinite-volume limit is taken on averages with respect to density matrices \eqref{rho}, in order to obtain averages of local observables in the thermodynamic limit. We denote the resulting infinite-volume averages by $\bra \cdots\ket$. In quantum chains with finite-dimensional local space, for instance, the infinite-volume limit is proven to exist and to give a unique state whenever $\beta^i Q_i$ (here and below, summation over repeated indices is implied unless otherwise stated) has finite-range or exponentially decaying interaction \cite{Araki}. Here we assume a state $\bra \cdots\ket$ is given, the thermodynamic limit having been taken.

Maximal entropy states of thermodynamic systems are expected to satisfy a number of crucial properties \cite{BratelliRobinson12}. The properties of interest for the state $\bra\cdots\ket$ involve the extensive conserved charges $Q_i$, as well as the particular charge
\beq\label{WQi}
	W = \beta^i Q_i
\eeq
which specifies the state. These concepts hold both for quantum and classical systems. 

The first required property is that the state be invariant under evolution with respect to all charges $Q_i$, and be translation invariant (homogeneous):
\beq\label{homo}
	\bra [Q_i,\Or]\ket = 0,\quad \nabla \bra  \Or(\v{x}) \ket = 0
\eeq
where again $\Or$ is any local observable or product thereof. The state is also clustering: connected correlation functions vanish at large distances,
\beq\label{clus}
	\bra \Or_1(\v{x})\Or_2(0)\ket^{\rm c} \to 0 \qquad (|x|\to\infty),
\eeq
where $\bra \Or_1(\v{x})\Or_2(0)\ket^{\rm c} = \bra \Or_1(\v{x})\Or_2(0)\ket- \bra \Or_1(\v{x})\ket\bra\Or_2(0)\ket$. For our purposes we require that the vanishing be at least integrable, i.e.~faster than $1/|\v{x}|^d$. In quantum spin chains, it is exponential \cite{Araki}.

Second, the Kubo-Martin-Schwinger (KMS) relation holds,
\beq \label{kms}
	\bra \Or_1\Or_2\ket = \bra \tau_{-\ri\hbar} \Or_2\,\Or_1\ket
\eeq
or equivalently 
\beq\label{kms2}
	\bra[\Or_1,\Or_2]\ket = \bra (1-\tau_{-\ri\hbar})\Or_1 \,\Or_2\ket.
\eeq
Here $\tau_t$ is the evolution with respect to the generator $W$,
\beq \label{tau_s}
	\tau_t \Or = e^{\ri t[W,\cdot]/\hbar}\Or.
\eeq
The KMS relation has an important role in the full characterisation of Gibbs states in the operator algebra formulation of quantum statistical mechanics \cite{IsraelConvexity,BratelliRobinson12}. It can be taken as a definition of the potentials $\{\beta^i\}$ via \eqref{WQi}, once a basis $\{Q_i\}$ is chosen.

Third, a ``tangent-manifold" relation holds, as was studied at length in \cite{Doyon2017}, see also  \cite{IsraelConvexity}. It states that derivatives of averages in the state $\bra\cdots\ket$ with respect to $\beta^i$ are given by connected correlation functions with the conserved quantity $Q_i$,
\beq\label{tangent}
	-\frc{\p}{\p\beta^i} \bra \Or\ket = \bra \Or Q_i\ket^{\rm c},
\eeq
where the derivative exists and is continuous in all $\beta^i$'s. The right hand side is $\bra \Or Q_i\ket^{\rm c} = \int \dd^d{x}\,\bra \Or q_i(\v{x})\ket^{\rm c}$, which can be seen to exist by the clustering property described above. This provides an alternative definition of the $\beta^i$'s, which has a clearer conceptual link to the exponential form of the GGE density matrix. Relations \eqref{kms} and \eqref{tangent} are related in that they give rise to consistent potentials $\{\beta^i\}$, see appendix \ref{appkmstangent}.

The main hypotheses on the state $\bra\cdots\ket$ are properties \eqref{homo}, \eqref{clus}, \eqref{kms} and \eqref{tangent}. Precise formulations of clustering and of the KMS relation require further considerations, see e.g.~\cite{Doyon2017} and \cite{BratelliRobinson12}. For our purposes, we assume that the KMS relation holds for local observables, and we further assume that clustering \eqref{clus} holds for imaginary-time evolved local observables: with $\Or_1(\v x)=\tau_{-\ri \hbar s}\t\Or_1(\v x)$ where $\t\Or_1$ is a local observable, uniformly for $s\in[0,1]$.

Finally, we make two additional assumptions. (1) The manifold of states is a connected submanifold $\mathcal M$ of the space of values of multiplets $\beta^\bullet$, and $\mathcal M$ contains the ``infinite-temperature state" $\beta^i=0\;\forall\;i$, which is assumed to have the trace property. In general $\mathcal M$ may be nontrivial, however in quantum spin chains, considering a finite number of local conserved charges, one can take all $\beta^i\in\R$, see appendix \ref{app_rigour} for a discussion. (2) The currents $\v{j}_{ij}$ as defined by \eqref{evol} are defined only up to the addition of terms proportional to the identity observable, and we make the assumption that it is possible to fix these constant terms so that all currents have vanishing averages in the infinite-temperature state -- this is the trace state in quantum lattice models. The gauge ambiguity of the currents is thus fixed by
\beq\label{gauge}
	\bra \v{j}_{ij}\ket_{(\beta^\bullet= 0)} =0.
\eeq 

Appendix \ref{app_rigour} contains a discussion of rigorous results on these properties for MES. In particular, it is shown that all assumptions and required structures are proven rigorously in quantum spin chains with finite-range interactions.

All hypotheses above are assumed to hold also in the classical realm. In this case, all observables are taken to be phase-space functions instead of operators, and one makes the replacements
\beq\label{poisson}
	\frc{\ri}{\hbar} [a,b] \rightarrow \{a,b\}\quad\mbox{and, in \eqref{kms2},}\quad
	(1-\tau_{-\ri\hbar})\Or_2\rightarrow\ri \hbar\{W,\Or_2\}
\eeq
where $\{\cdot,\cdot\}$ is the Poisson bracket. The classical KMS relation is equivalent to relation \eqref{kms2} to order $\hbar$; this relation can be shown to fully characterise the Gibbs states \cite{aizenman1977equivalence}. From the viewpoint of the operator algebra in the quantum (classical) case, observables form a non-commuting (commuting) algebra, and one uses commutators (Poisson bracket), as per Eq.~\eqref{poisson}.

For simplicity, in the subsequent calculations we consider only quantum systems, and set $\hbar=k_B=1$. 

\subsection{Free energy fluxes and entropy currents}\label{ssectfluxes}

A number of fundamental relations hold for the averages of currents and densities in a MES, see for instance \cite{Spohn-book}. In fact, many of these relations hold more generally in homogeneous, $Q_k$-invariant, clustering states.

One of the most important is a symmetry relation involving correlation functions of currents and densities. We consider the integrated density-charge correlator
\beq \label{B_mat}
	\v B_{kij} = \int \dd^d{x}\,\bra \v{j}_{ki}(0) q_{j}(\v{x})\ket^{\rm c} = -\frc{\p}{\p\beta^j} \bra \v{j}_{ki}\ket,
\eeq
from which we form the set of vector-valued matrices $\v B_{k\bullet\bullet}$, where $k$ is the index of the charge generating the currents. These matrices are {\em symmetric} in the last two indices,
\beq \label{B_symm_eq}
	\v B_{kij} = \v B_{kji}.
\eeq
In one dimension of space such a symmetry was shown in a family of classical, stochastic, interacting particle systems in \cite{toth2003onsager,grisi2011current}. There, the importance of this Onsager-type relation for the physical consistency of the emerging hyperbolic system of hydrodynamic equations was pointed out. In the present context, Eq.~\eqref{B_symm_eq} was shown in \cite{PhysRevX.6.041065,dNBD2} for one dimensional systems. The extension of such derivations to higher dimensions is immediate. The most general relations of this type were obtained in \cite{KarevskiCharge2019}, where in particular the assumption of space-translation invariance is relaxed. See also \cite{Spohn-book,SpohnNonlinear,SciPostPhys.3.6.039,DoyonLecture2020} for discussions of hydrodynamic matrices. For completeness we provide a simple derivation of \eqref{B_symm_eq} in appendix \ref{app_symmetry_B_mat}. 

This symmetry implies that there must exist, for every $k$,  a differentiable function $\v{g}_k$ from $\mathcal M$ to $\R^d$ which generates the currents according to \cite{PhysRevX.6.041065}, 
\beq \label{free_flux}
	\bra \v{j}_{ki}\ket = \frc{\p \v{g}_k}{\p\beta^i}.
\eeq
This parallels the situation for the specific free energy $f$, where the more apparent symmetry of the static covariance matrix $\mathsf C_{ij} =\mathsf C_{ji}= \int \dd^d{x}\,\bra q_{i}(0) q_{j}(\v{x})\ket^{\rm c} = -\frc{\p}{\p\beta^j} \bra q_{i}\ket$ implies that there exists\footnote{In this paper, we define the specific free energy by absorbing the temperature factor that usually appears in its definition; this is more natural, as we consider all charges on equal footing.} $f$ such that $\bra q_{i}\ket = \frc{\p}{\p\beta^i} f$. The quantities $\v{g}_k$ are referred to as ``free energy fluxes".

The free energy fluxes appear in the universal definition of the entropy current. That the symmetry \eqref{B_symm_eq} implies the existence of a Lax entropy (see e.g.~\cite{BressanNotes}) for the hydrodynamic equations was first noticed in \cite{toth2003onsager,grisi2011current}, see \cite{DoyonLecture2020} for a discussion in the present context.  The entropy current $\v{\mathsf j}^{ s}_k$ with respect to the flow generated by $Q_k$ is defined as
\beq\label{sk}
	\v{\mathsf j}_k^{s} = \beta^j \bra\v{j}_{kj}\ket - \v{g}_k.
\eeq
This parallels the definition of the usual entropy density
\beq\label{s}
	s = \beta^j \bra q_j\ket - f.
\eeq
Again, for the physical entropy current generated by the Hamiltonian we have:
\beq
	\v{\mathsf j}^{s}= \v{\mathsf j}_2^{s}\qquad\mbox{(physical entropy current).}
\eeq
The physical meaning of the entropy currents and density is clearest when considering the hydrodynamic equations that arise at the Euler scale, see subsection \ref{ssecteuler}.

Recalling our choice of the currents with respect to space translation, Eq.~\eqref{j1i}, using \eqref{free_flux} we identify the free energy flux with respect to the momentum as the free energy (if a conserved momentum is present),
\beq\label{g1}
	\v{g}_{1_\alpha} = \v{e}_{\alpha} f \qquad \mbox{(recall that $Q_{1_\alpha}=P_\alpha$ is the total momentum).}
\eeq
In particular, combining with \eqref{j1i}, this shows that $\v {\mathsf j}^{s}_{1_\alpha} = \v{e}_\alpha s$.

\section{Main results: an Euler-scale KMS relation}\label{sectMain}

The goal of this paper is to obtain a general relation for the free energy fluxes in a MES, which follows from the KMS relation. This general relation states that the contraction of the free energy fluxes onto the potentials is a state-independent spatial vector:
\beq\label{gkms1}
	\beta^k \v{g}_k = \v{G},
\eeq
with summation over repeated indices. In other words, for every spatial component $\alpha$, the charge vector $g_\bullet^\alpha$ must lie on a plane perpendicular to $\beta^\bullet$. Equation \eqref{gkms1}, and its corollaries \eqref{gkmsent} \eqref{Gentropy} and \eqref{gkmspar} below, are our main results. As \eqref{gkms1} is a consequence of the KMS equation as applied to the currents, which are the crucial ingredients for Euler-scale hydrodynamics, we will refer to it as the Euler-scale KMS (EKMS) relation.  We show relation \eqref{gkms1} in section \ref{sectproof} under the hypotheses of subsection \ref{ssectMES}, based on an important relation that we establish in section \ref{sectmoment}. In appendix \ref{app_rigour}, it is explained how the proof of \eqref{gkms1} is mathematically rigorous in quantum spin chains; Theorem \ref{theochain} expresses the result accurately.

A further general result is that the constant spatial vector $\v{G}$ determining the plane also satisfies (recall the entropy currents $\v{\mathsf j}^{(s)}$ defined in \eqref{sk}), 
\beq\label{gkmsent}
	\beta^k \v{\mathsf j}^{s}_k = -2\v G.
\eeq
That is, the entropy current for evolution with respect to the charge $W=\beta^k Q_k$ defining the state, takes a constant, state-independent value. Writing this relation in a thermal state, we have
\beq\label{Gentropy}
	\v{G} = -\frc1{2T} \v{\mathsf j}^{s}\qquad \mbox{(in a thermal state at temperature $T$, that is, $W =  H/T$).}
\eeq
This shows that the constant $\v G$ is fixed by the physical entropy current. A trivial implication is thus that the physical entropy current is proportional to the temperature in a thermal state.

However, the physical entropy current is usually vanishing at equilibrium, as a consequence of parity symmetry. More generally, suppose that for every spatial component $\alpha$, there exists at least one conserved quantity $Q_i$ that is $x^\alpha$-parity symmetric -- invariant under $x^\alpha\mapsto -x^\alpha$ -- and that is ``stable" -- such that a configuration with $\beta^i\neq 0,\,\beta^{j}=0 \, (j\neq i)$ is a well defined state (lies in the manifold of states $\mathcal M$). Then the constant $\v{G}$, hence in particular the thermal entropy current, vanishes,
\beq\label{gkmspar} 
	\v{G} = 0 \qquad\mbox{(with parity symmetry).}
\eeq
For instance, this is the case if the Hamiltonian itself, $H=Q_2$, is parity symmetric for all directions. Importantly, it is not necessary for the Hamiltonian itself to be parity symmetric in order for \eqref{gkmspar} to hold, as long as,  for every spatial direction, there exists a conserved charge as above which is parity symmetric for this direction.

Relations \eqref{gkmsent} and \eqref{gkmspar} are shown below to follow from \eqref{gkms1}. Some additional relations are of interest. Differentiating \eqref{gkms1} with respect to $\beta^i$, we obtain a relation between the free energy fluxes and the average currents:
\beq\label{gibeta}
	\v{g}_i = -\beta^k \bra \v{j}_{ki}\ket.
\eeq
Thus, the free energy fluxes are expressible in terms of averages of local observables. This generalises the well-known thermodynamic relation between the specific free energy $f$ and the pressure, or average momentum current, in thermal states. Recall the stress tensor \eqref{stress}, in our notation ${\cal T}^\gamma_\alpha = j^\gamma_{1_\alpha}$. Taking \eqref{gibeta} with respect to the momentum flow $i=1_\alpha$ in a thermal state with temperature $T=1/\beta^2$ ($\beta^k=0$ for $k\neq 2$),
with \eqref{g1} we arrive at
\beq\label{fT}
	f = -\frc1T \bra {\cal T}^\alpha_\alpha\ket \qquad\mbox{(no summation over $\alpha$).}
\eeq
This is the standard relation between the specific free energy and pressure. In particular, this shows that $\bra {\cal T}^\alpha_\alpha\ket $ is isotropic (independent of the direction $\alpha$) in any thermal state of a system with short-range interactions.

Relation \eqref{gibeta} allows us to express all entropy currents solely in terms of local averages. We define the state-dependent local observable
\beq
	\v{j}_k^{s} = \beta^j (\v{j}_{kj} + \v{j}_{jk}),
\eeq
so that the entropy currents are
\beq
	\v{\mathsf j}_k^{s} = \bra\v{j}_k^{s}\ket.
\eeq
This generalises the thermodynamic relation $s=\bra \mathcal T_0^0+ \mathcal T_\alpha^\alpha\ket/T$, in thermal states of temperature $T$, which relates the entropy density $s$ to the expectation of the energy density $\mathcal T_0^0=q_2$ and of the stress $\mathcal T_\alpha^\alpha$ (no summation over $\alpha$).

Finally, differentiating \eqref{gibeta} we obtain a symmetry relation for the indices of average currents:
\beq \label{swap_indices}
	\bra \v{j}_{ij} + \v{j}_{ji}\ket = -\beta^k \frc{\p}{\p\beta^j}\bra \v{j}_{ki}\ket = \beta^k \v{B}_{kij},
\eeq
and the related equations:
\beq\label{betabetaj}
	\beta^k\beta^i \bra \v{j}_{ki}\ket =\frac{1}{2}\beta^k\beta^i\beta^j \v B_{kij} = -\v{G}.
\eeq

{\em How to obtain \eqref{gkmsent}:} We assume \eqref{gkms1}, thus \eqref{gibeta} holds. Eq.~\eqref{gkmsent} then immediately follows from the definition \eqref{sk}.

{\em How to obtain \eqref{gkmspar}:} Again we assume \eqref{gkms1}, thus \eqref{gibeta} holds. Suppose that one of the charges, say $Q_i$ for some given $i$, is $x^\alpha$-parity symmetric such that there exists a parity transformation $\mathcal P$ with $\mathcal P(q_i(\v{x})) = q_i(x^1,\ldots,-x^{\alpha},\ldots,x^d)$. Suppose that we may specialise relation \eqref{gibeta} to the MES determined by $\beta^j=0\,(j\neq i)$, that is $W = \beta^iQ_i$ (no summation), giving $\v{g}_i = -\beta^i \bra \v{j}_{ii}\ket$ (no summation). The state is $x^\alpha$-parity symmetric, and by the continuity equations \eqref{evol}, the current $\v{j}_{ii}$ transforms as
\[
	\mathcal P(j_{ii}^\gamma(\v{x})) = (-1)^{\delta_{\alpha,\gamma}}j_{ii}^\gamma(x^1,\ldots,-x^{\alpha},\ldots,x^d) + a
\]
where $a$ is a local observable that does not depend on position. As $a$ is independent of position, by clustering, $\bra a \Or(x)\ket^{\rm c}$ = 0, whence by \eqref{tangent} $\bra a\ket = c$ is independent of the state. Therefore, we find that $\bra j_{ii}^\alpha\ket = c/2$ is independent of $\beta^i$. By the choice of gauge \eqref{gauge} we therefore have $\bra j_{ii}^\alpha\ket = 0$. Thus, in this state, we may set $g_i^\alpha=0$. As we must have $G^\alpha = \beta^i g_i^\alpha$ (no summation), we deduce that the state-independent constant $G^\alpha$ vanishes. This gives \eqref{gkmspar}.

\section{Applications and examples}\label{sectAppl}

In this section, we provide applications and examples of the main relation \eqref{gkms1}. We show how it constrains the equations of state of systems of conventional type, which have only a few conserved quantities, and the thermodynamic Bethe ansatz structure of integrable systems, which have an infinite number of conserved quantities. We also explain how it implies that stationary solutions of the Euler hydrodynamic equations  \eqref{eulerpotential} within external potentials have the correct form, and that these equations preserve entropy. The latter puts on a common framework the standard expectation for conventional hydrodynamics, and the results already established in the generalised hydrodynamics of integrable systems \cite{DY}. Finally, we show how the main result holds in the explicit example of $d$-dimensional conformal hydrodynamics, where the free energy fluxes were evaluated in \cite{DoMy19}. In this section we do not aim for mathematical rigour.

\subsection{Systems with few conserved quantities}\label{ssectfew}
Consider the manifold of states $\mathcal M$ built from only the particle number $N=Q_0$ (if admitted by the system), the total momentum $P_\alpha = Q_{1_\alpha}$ and the energy $H=Q_2$, and no other conserved charge (that is, $\beta^i=0$ for all other indices $i$). Typical conventional non-integrable systems admit only these as conserved charges. Here we interpret $N$ physically as a particle number, and hence its density is ``ultra local": it does not involve any coupling between points in space. As a consequence, the flow generated by $N$ has trivial currents $\v{j}_{0i}=0$, and therefore we take $\v g_0=0$. This can be taken as the definition of $Q_0$ as a conserved quantity with an ultra-local density. We denote $\beta^2 = \beta=1/T$ (the inverse temperature), $\beta^{1_\alpha}=-\beta\nu^\alpha$ ($\nu^\alpha$ is a parametrisation of the boost, if Galilean or relativistic invariance is present), and $\beta^0 = -\beta \mu$ ($\mu$ is the chemical potential, if there is particle number conservation). The  density matrix for the Gibbs state is of the form
\beq
	\rho\propto e^{-\beta(H-\v{\nu}\cdot\v{P}-\mu N)}.
\eeq
The results \eqref{g1} and \eqref{gkms1} then imply
\beq
	\v{g}_2 = T\v{G}+\v{\nu} f.
\eeq
Recall that, by \eqref{Gentropy}, the quantity $-2\v{G}T$ is the entropy current $\v{\mathsf j}^{s}$ in an equilibrium thermal state at temperature $T$. Again, if for instance the Hamiltonian is parity symmetric, then this current vanishes, $\v{G}=0$. Evaluating the currents as $\bra \v{j}_i\ket = \p \v{g}_2/\p\beta^i$, we obtain, for the average currents of particles $\bra \v j_0\ket$, of the direction-$\alpha$ momentum component $\bra \mathcal T_\alpha^\gamma\ket$, and of the energy $\bra \v j_2\ket$, the expressions
\beq\label{currentfew}
    \bra\v{j}_0\ket=\v{\nu} \bra q_0\ket,\quad
    \bra\mathcal T_\alpha^\gamma \ket = \nu^\gamma\bra q_{1_\alpha}\ket -Tf\delta_\alpha^\gamma,\quad
    \bra \v{j}_2\ket = \v \nu \bra q_2\ket - T\v \nu f-T^2 \v{G} .
\eeq
Knowledge of the free energy $f$, or equivalently of the static covariance matrix $\mathsf C_{ij}$, would allow us to write $\v\nu, T$ in terms of the average particle density $\bra q_0\ket$, direction-$\alpha$ momentum density component $\bra q_{1_\alpha}\ket$, and energy density  $\bra q_2\ket$. Therefore, {\em the equations of state are fully fixed by the free energy $f$} (and the constant vector $\v G$ characterising the thermal entropy current, which vanishes if the Hamiltonian is parity symmetric). In Galilean or relativistically invariant systems, space-time symmetries can be used to establish the form \eqref{currentfew} of the currents. 
Here however, we find that the currents are fixed in this manner solely as a consequence of the KMS relation, independently of the presence or otherwise of any space-time boost symmetries.
%

\subsection{Integrable Systems}
In one-dimensional ($d=1$) integrable models, an explicit expression for the free energy fluxes is known within the framework of the thermodynamic Bethe ansatz (TBA) \cite{PhysRevX.6.041065,PhysRevLett.117.207201}. For simplicity, let us assume that the TBA formulation involves only one quasiparticle type, with rapidities ranging over $\R$. This covers many models, for instance in the Lieb-Liniger model of quantum gases \cite{Yang-Yang-1969} and the Toda model of classical particles (see e.g.~\cite{SpoToda,DToda,BCM19} and references therein) -- it is simple to extend the discussion to more complicated spectra.

The free energy fluxes take the form \cite{PhysRevX.6.041065}
\beq
	g_k = \int \dd \theta\,\p_\theta h_k(\theta)\, F(\ep(\theta)),
\eeq
where $h_k(\theta)$ is the one-particle eigenvalue of $Q_k$ on a state with a quasiparticle of rapidity $\theta$, and $F(\ep)$ is the free energy function of the pseudo-energy $\ep(\theta)$. For example, in fermionic integrable models $F(\ep)=-\log(1+e^{-\ep})$ and in classical particle systems $F(\ep) = -e^{-\ep}$, see e.g.~\cite{DoyonLecture2020}. Generally the pseudo-energy solves the equation:
\begin{equation}
    \ep(\theta)=\sum_i\beta^ih_i(\theta)+\int d\theta'\,\varphi(\theta',\theta)F(\ep(\theta'))
\end{equation}
with $\varphi(\theta',\theta)$ the two-particle differential scattering phase of the model. Further, we assume that
\beq\label{asym}
	\int_{\ep(-\infty)}^{\ep(\infty)}\dd \ep \,F(\ep) = 0.
\eeq
A sufficient condition for this to hold is that $\lim_{\theta\to\pm\infty}\ep(\theta) = \infty$, and $\lim_{\ep\to\infty}F(\ep)=0$ (sufficiently fast). The former condition is seen to hold in many physically relevant states (see e.g.~\cite{Yang-Yang-1969,ZAMOLODCHIKOV1990695,TakahashiTBAbook}), and the latter is immediate with the free energy functions above. 

The relation \eqref{gkms1} then implies
\beqa
	\beta^kg_k &=&\sum_k\beta^k \int \dd \theta\,\p_\theta h_k(\theta)\, F(\ep(\theta)) \n
	&=& \int \dd \theta\,\Big(\p_\theta\ep(\theta)+\int \dd\theta'\, \p_\theta\varphi(\theta',\theta) F(\ep(\theta'))\Big) F(\ep(\theta)) \n	
	&=& \int \dd \theta\int \dd\theta'\, \p_\theta\varphi(\theta',\theta) F(\ep(\theta'))F(\ep(\theta)).
	\label{betagF}
\eeqa
Since $\varphi(\theta',\theta)  = \p_{\theta'} \phi(\theta',\theta)$, where $\phi(\theta',\theta)=-\ri\,\log S(\theta',\theta)$ is the complex phase of the two-body scattering amplitude $S(\theta',\theta)$, the requirement that \eqref{betagF} vanish in arbitrary states is equivalent to the condition
\beq
	\p_{\theta'}\p_\theta \phi(\theta',\theta) = - 
	\p_{\theta'}\p_\theta \phi(\theta,\theta').
\eeq

Thus the EKMS relation implies that the scattering phase $\phi(\theta',\theta)$ be anti-symmetric, up to ``factorised" terms of the type $u(\theta') + v(\theta)$. Anti-symmetry of $\phi(\theta',\theta)$ is, in quantum systems, the condition of unitarity of the two-body scattering amplitude, and therefore we find that the EKMS relation implies unitarity of the TBA scattering amplitude up to these factorised terms. The conclusion holds also for classical systems. Factorised terms would deserve further studies, although they likely may be argued to vanish by appealing to other physical properties of the scattering amplitude; for instance, one generally expects that the scattering amplitude tends to a constant at large rapidity.

\subsection{Euler-scale hydrodynamic equations}\label{ssecteuler}

The Euler-scale hydrodynamic equations constitute one of the most important applications of the expressions of currents in MES. In this subsection, we obtain general results on Euler-scale hydrodynamics that follow from the EKMS relation.

Euler-scale hydrodynamic equations are hard to show rigorously, and we do not make any attempt at rigour here. For completeness, we nevertheless present the heuristic arguments, based on local relaxation, underlying the universal form of the Euler-scale hydrodynamic equations with external force coupled to arbitrary charges. Explicit forms of Euler-scale hydrodynamic equations depend on the explicit expressions of currents in MES, which is also in general a hard problem (but see subsection \ref{ssectfew}). Here we consider the universal form, which depends abstractly on the average currents, without having to evaluate them.

The Euler equations are based on the assumption of local entropy maximisation \cite{Spohn-book} with respect to the time-evolution generator $U$. In our discussion above we took $U$ to be the Hamiltonian. Let us here include more generally space-varying external potentials $u^k(\v x)$,
\beq
	U=\int \dd^d x\,u^k(\v x)q_k(\v x).
\eeq
For example, we could take $U = H + V$, where $V$ represents a coupling to external spatially-varying fields $V = \int \dd^d x\,u^0(\v x)q_0(\v x)$ (in the notation of subsection \ref{ssectfew}); or a spatially-varying Hamiltonian strength $U=\int \dd^d x\,u^2(\v x)q_2(\v x)$.

Assuming local entropy maximisation, averages of time-evolved local observables, $\Or(\v{x},t) = e^{\ri Ut}\Or(\v{x})e^{-\ri Ut}$, in non-homogeneous initial states $\bra\cdots\ket_{\rm ini}$, are well described by space-time dependent MES $\bra\cdots\ket_{\v{x},t}$, that is $\bra \Or(\v{x},t)\ket_{\rm ini} \approx \bra\Or\ket_{\v{x},t}$. Physically, $\bra\cdots\ket_{\v x,t}$ is the state in the local fluid cell at $\v x,t$, according to the hydrodynamic separation of scales. The dynamics of the space-time dependent state are fixed by the conservation laws, yielding the Euler equations of the system. Below we drop the indices $\v x,t$ for lightness of notation, and $\bra\cdots\ket$ represents the local state in the fluid cell.

Local entropy maximisation is expected to hold approximately only under certain conditions; such that the initial state $\bra\cdots\ket_{\rm ini}$ and potentials $u^k(x)$ vary only at long wavelengths, and the time $t$ is large enough for local relaxation to have occurred.

From the assumption of local entropy maximisation, a heuristic derivation of the Euler equations follows as in \cite{DY}. The Heisenberg equation of motion (with the replacement \eqref{poisson} for classical systems) for the charge densities reads:
\beq
    \p_tq_i(\v{x})-\ri\int \dd^d{y}\,u^k(\v{y})[q_k(\v{y}),q_i(\v{x})]=0.
\eeq
We now assume that $u^k(\v{x})$ varies only at long wavelengths, and take a derivative expansion,
\beq
    \p_tq_i(\v{x})-\ri\int \dd^d{y}\,\left(u^k(\v{x})+(\v{y}-\v{x})\cdot\nabla u^k(\v{x})\right)[q_k(\v{y}),q_i(\v{x})]=O(\nabla^2).
\eeq
The Euler-scale hydrodynamic equation is obtained by taking averages, and making the local entropy maximisation assumption at the first-derivative order. By \eqref{evol} the first term in the parenthesis gives, after integration over $\v y$ and averaging, the divergence of the current $u^k\nabla \cdot \bra \v{j}_{ki}\ket$; recall that $\bra\cdots\ket$ is the MES of the local fluid cell. For the second term, as the coefficient $\nabla u^k$ already contains a derivative, the integral over $\v y$ may be assumed to lie entirely within the local, homogeneous fluid cell, $\int \dd^d y\, (\v y-\v x)\cdot \bra [q_k(\v y),q_i(\v x)]\ket$. Using the result \eqref{firstmoment} of section \ref{sectmoment},
we then obtain
\begin{equation}\label{eulerpotential}
    \p_t\bra q_i\ket +\nabla\cdot(u^k\bra\v{j}_{ki}\ket)+\nabla u^k\cdot \bra\v{j}_{ik}\ket=0.
\end{equation}

This is the universal form of the Euler equation for the many-body system. As far as we are aware, it was first written in \cite{DY} (where the relation \eqref{firstmoment} was argued for under sightly stronger assumptions than those made here). Importantly, the total charges $Q_i$ are no longer conserved in spatially-varying external potentials, as the extra term $\nabla u^k\cdot \bra\v{j}_{ik}\ket$ is not a total derivative; however it is simple to see that the inhomogeneous operator $U$ is conserved. Thus, if the homogeneous model is integrable, the inhomogeneous time-evolution operator breaks integrability. In the context of integrable systems, an extension of this equation to the diffusive order, with certain large-wavelength external potentials, was studied in \cite{bastianello2020thermalisation}.


In the special case where the $u^k$ are position-independent the equation reduces to
\begin{equation}\label{euler}
    \p_t\bra q_i\ket +\nabla\cdot(u^k \bra\v{j}_{ki}\ket)=0.
\end{equation}
In this case all charges are conserved, as the evolution is generated by the conserved charge $U=u^k Q_k$. We note that $u^k\v j_{ki}$ is the $i^{\rm th}$ current with respect to evolution generated by the charge $U$. At space-time points where the solution is smooth, it is straightforward to show (see e.g. \cite{toth2003onsager,grisi2011current,DoyonLecture2020}) that this equation implies conservation of entropy,
\beq
	\p_{t} s + \nabla\cdot (u^k\v{\mathsf j}_k^{s}) = 0,
\eeq
where we use \eqref{sk} and \eqref{s}. The quantity $s$ is a Lax entropy for the hyperbolic system \eqref{euler} \cite{toth2003onsager,grisi2011current,BressanNotes}. Thus, $u^k\v{\mathsf j}_k^{s}$ is interpreted as the entropy current with respect to the evolution generated by the charge $U$.


Using \eqref{gkms1}, we obtain two general results under \eqref{eulerpotential}: 
\bi
\item[(i)] At every point where the solution is smooth, the entropy density satisfies an exact continuity equation, and thus the total entropy is conserved except possibly at non-smooth points.
\item[(ii)] Under mild genericity assumptions, the space-dependent MES $\bra\cdots\ket_{\v x}$ which is a stationary solution of \eqref{eulerpotential}, can be seen as the fluid cell approximation (local density approximation) of a thermal state with respect to the space-varying generator $U$. Its formal density function is $\rho_{\v x}\propto e^{-\b \beta (u^k(\v x)Q_k- \b\mu Q_0)}$ for some inverse temperature $\b \beta$ and (if there is particle conservation $Q_0$) chemical potential $\b \mu$, both independent of $\v x$.
\ei
The first result is natural at the Euler scale; despite total charges not being conserved in external inhomogeneous fields, one expects entropy to still be preserved at this scale by virtue of the lack of any diffusive effects and the local maximisation of entropy\footnote{An exception is made for instance at shocks (weak solutions to the Euler equation).}. This is obtained here in full generality with arbitrary, long-wavelength inhomogeneous external fields. The second result suggests that if a weak irreversibility is added to the Euler equation, the stationary solution approached must be thermal. These solutions are typically inaccessible by the Euler dynamics due to the entropy reduction required to access them, however they are approached if diffusive effects are added \cite{bastianello2020thermalisation}.

In order to show statement (i), we consider the definitions \eqref{sk} and \eqref{s} of the entropy density and currents. Applying the time derivative to \eqref{s} we find
\beq
	\p_t s = \beta^i \p_t \bra q_i\ket =  -\beta^i\big(\nabla\cdot( u^k \bra\v j_{ki}\ket) +\nabla u^k \cdot \bra\v j_{ik}\ket\big).
\eeq
On the other hand, the divergence of the entropy current $u^k \v{\mathsf j}_{k}^{s}$ generated by $U$ is
\beq
	\nabla\cdot \big(\beta^i u^k \bra\v j_{ki}\ket - u^k \v g_k\big) =
	\beta^i \nabla\cdot(u^k \bra\v j_{ki}\ket) - \nabla u^k\cdot \v g_k.
\eeq
Using \eqref{gibeta} in the form $\v g_k = -\beta^i \bra \v j_{ik}\ket$, we obtain the required cancellation
\beq
	\p_t s + \nabla\cdot (u^k \v{\mathsf j}_k^{s}) = 0.
\eeq
In order to show statement (ii), we use the chain rule and the identity \eqref{swap_indices} for the $\v B$ matrix. The condition for stationarity $\p_t\bra q_i\ket=0$ of \eqref{eulerpotential} then reads
\begin{equation}
    \left(u^k\nabla\beta^j-\beta^k\nabla u^j\right)\cdot\v B_{kij}=0.
\end{equation}
As this must hold for all $i$, generically, it must be true for each ordered pair $(j,k)$ separately; and if the $\v B$-matrices are generic enough, the parenthesis must vanish when projected onto a generic vector. Thus, we have the solution
\begin{equation}\label{betak}
    \beta^k=\b\beta u^k,
\end{equation}
with $\b\beta$ a $k$-independent thermalisation constant. This is the locally thermal state with respect to the generator $U$, and indicates that the density matrix of the stationary state is of the form $\rho\propto e^{-\b\beta U}$. Thus $\b\beta$ is identified with the inverse temperature of the stationary state. 

We have used arguments of genericity. Of course, if the $\v B$ matrices satisfy certain relations, or do not span a large enough vector space, then different, special solutions may exist; it would be interesting to investigate these solutions and their physical meaning. A case which is important is that where an ultra-local conserved density exist, such as from particle conservation, $N=Q_0$. As mentioned, this means $\v{g}_0=0$, hence $\v{B}_{0ij}=0$ for all $i,j$. For $\beta^0$, the generic stationarity requirement becomes, instead of \eqref{betak},
\begin{equation}
    \beta^0=\b\beta (u^0-\b \mu),
\end{equation}
where an additional constant -- the chemical potential of the stationary state -- appears. In this case, the stationary density matrix  is of the form $\rho\propto e^{-\b\beta (U-\b \mu Q_0)}$.

We note that the statements (i) and (ii) were shown in \cite{DY} in the context of integrable models, by using the specific properties of generalised hydrodynamics. Here they are shown to arise from fully general principles of many-body physics. One expects corresponding statements (increase of entropy, thermal stationary state) for the hydrodynamic equation that includes the diffusive corrections. With ultra-local space-varying external field, such statements were shown in \cite{bastianello2020thermalisation} in the context of integrable models. The EKMS relation can likewise be used to obtain these in the more general context of many-component hydrodynamic equations, without the use of integrability. We plan on presenting a more extensive discussion of this and related aspects in a future work.

\subsection{Conformal hydrodynamics}

The result \eqref{gkmspar} can be shown explicitly in conformal hydrodynamics. Consider a Lorentz-invariant conformal field theory in $d>1$ spatial dimensions. The MES's of this model are the boosted thermal states, where we can use the rotational symmetry of the theory to consider a boost only along the $x^1$-direction (with $x^0=t$), thus involving only two potentials\footnote{Connecting to the general notation introduced in \eqref{gennot}, we have $\beta^1 = \beta^{1_1}$.} $\beta^i$ for $i=1,2$:
\begin{equation}\label{conformal_state}
    \rho\propto e^{-\beta^1P_1-\beta^2H}=e^{-\beta_{\rm rest}(\mathrm{cosh}(\theta)H-\sinh(\theta)P_1)}.
\end{equation}
Here $\beta_{\rm rest}$ is the inverse temperature in the rest-frame, and $\theta$ is the rapidity of the boost. In terms of the energy-momentum tensor $\mathcal T^{\mu\nu}$, we have $H=Q_2=\int d^d\v{x}\, \mathcal T^{00}(\v{x})$ and $P_1=Q_1=\int d^d\v{x}\, \mathcal T^{01}(\v{x})$. The charge and current averages within the state \eqref{conformal_state} are constrained as follows by the conformal symmetry:
\beq
	\bra \mathcal T^{\mu\nu}\ket = a \beta_{\rm rest}^{-d-1}((d+1)u^\mu u^\nu +\eta^{\mu\nu}),\quad
	\eta^{\mu\nu} = {\rm diag}(-1,1,1,\ldots,1)
\eeq
where $a$ is a model-dependent constant and $u^\mu = (\cosh\theta,\sinh\theta,0,0,\ldots,0)^\mu$. In particular,
\begin{equation}
    \bra q_1\ket=\bra j_2\ket=a(d+1)\beta^{-(d+1)}_{\rm rest}\mathrm{cosh}(\theta)\mathrm{sinh}(\theta),
\end{equation}
\begin{equation}
    \bra q_2\ket=a\beta^{-(d+1)}_{\rm rest}\left(d\,\mathrm{cosh}^2(\theta)+\mathrm{sinh}^2(\theta)\right),
\end{equation}
\begin{equation}
    \bra j_1 \ket=a\beta^{-(d+1)}_{\rm rest}\left(\mathrm{cosh}^2(\theta)+d\,\mathrm{sinh}^2(\theta)\right).
\end{equation}
%
%

We would like to verify the general relation \eqref{gkms1} in this model of hydrodynamics. As parity symmetry is generically present, we have $\v G=0$. As we only have two potentials, we concentrate on the free energy fluxes $g_1$ and $g_2$. It turns out that these have been evaluated in \cite{DoMy19}, and take the form
\begin{equation}
    g_1=f=-\beta_{\rm rest}^{-d}\mathrm{cosh}(\theta),\quad
    g_2=-\beta_{\rm rest}^{-d}\mathrm{sinh}(\theta).
\end{equation}
It is a simple matter to verify that the EKMS \eqref{gkms1} with $G=0$, relation $\beta^1g_1 + \beta^2 g_2=0$ is indeed satisfied. In particular, it is clear in this example that the EKMS relation implies that the knowledge of $f$ immediate gives $g$, hence the average currents, without explicitly using relativistic and scale invariance. This is a special case of the situation considered in section \ref{ssectfew}.

\section{A first-moment relation}\label{sectmoment}

In this section we establish, under the hypotheses of subsection \ref{ssectMES}, a general identity for the symmetric sum of current averages,
\beq\label{firstmoment}
	\bra \v{j}_{ij} + \v{j}_{ji}\ket = -\ri \int \dd^d x\,\v{x} \bra[q_i(\v{x}),q_j(0)]\ket.
\eeq
In the classical case, using \eqref{poisson}, this is
\beq
	\bra \v{j}_{ij} + \v{j}_{ji}\ket = - \int \dd^d x\,\v{x} \bra\{q_i(\v{x}),q_j(0)\}\ket \quad\mbox{(classical case)}.
\eeq

This relation has previously been obtained by an assumed derivative expansion of the charge density commutator appearing on the right hand side \cite{DY}. Here we show that the result is valid under the very general conditions expressed in subsection \ref{ssectMES}. Relation \eqref{firstmoment} was used in \cite{DY} (as recalled in subsection \ref{ssecteuler}) in order to derive, from the assumption of local entropy maximisation, the general Euler-scale hydrodynamic equation in the presence of inhomogeneous external fields, Eq.~\eqref{eulerpotential}. It is also used in section \ref{sectproof} in order to establish our main result, the EKMS relation \eqref{gkms1}.
In this and the next section, the calculation for classical systems proceeds almost identically to the one presented here for quantum systems, and will be omitted.

We first define the marginals of the current operators by integrating over the transverse spatial degrees of freedom:
\beq \label{transverse}
    j_{ij,\alpha}(x^\alpha)=\int\prod_{\gamma\neq\alpha} \dd x^\gamma\,j^\alpha_{ij}(\v{x})
\eeq
Here, the upper $\alpha$ index refers to the vector component (we take the $\alpha$ on the left hand side as a subscript to indicate that the quantity is not actually a vector from the viewpoint of spatial transformations). In this notation, we have for some local observable $\Or$ at any given position,
\beqa
	\bra j_{ij,\alpha}(0)\Or\ket^{\rm c} &=& \int \prod_{\gamma\neq\alpha} \dd x^\gamma \int_{-\infty}^0 \dd x^\alpha\,\bra \p_{x^\alpha} j_{ij}^\alpha(\v{x})\Or\ket^{\rm c}\quad \mbox{(no summation over $\alpha$)}\n
	&=& -\ri\int\prod_{\gamma\neq\alpha} \dd x^\gamma\int_{-\infty}^0\dd x^\alpha\,\bra [Q_i, q_{j}(\v x)]\Or\ket^{\rm c}\n
	&=& -\ri\int \dd^d y \int\prod_{\gamma\neq\alpha} \dd x^\gamma \int_{-\infty}^0 \dd x^\alpha \,\bra [q_i(\v y), q_{j}(\v x)]\Or\ket^{\rm c}\n
	&=:& -\ri \int_{-\infty}^\infty \dd y\int_{-\infty}^0 \dd x\,\bra [q_{i,\alpha}(y), q_{j,\alpha}(x)]\Or\ket^{\rm c}.\label{step11}
\eeqa
The first equality holds by the clustering property \eqref{clus}  (in particular, clustering faster then $1/|\v x|^d$ is sufficient), and in the second we used the conservation laws \eqref{evol}. In the final equality we have defined $q_{i,\alpha}(x)$ analogously to \eqref{transverse}. Therefore
\begin{align}
	\bra (j_{ij,\alpha}(0)+& j_{ji,\alpha}(0))\, \Or\ket^{\rm c} = \\
	&
	-\ri \int \dd y\int_{-\infty}^0 \dd x\,\bra \big([q_{i,\alpha}(y), q_{j,\alpha}(x)] + [q_{j,\alpha}(y), q_{i,\alpha}(x)]\big)\Or\ket^{\rm c}\n
	=&
	-\ri \int \dd y \int_{-\infty}^{-y} \dd x\,\bra \big([q_{i,\alpha}(y), q_{j,\alpha}(x+y)] + [q_{j,\alpha}(y), q_{i,\alpha}(x+y)]\big)\Or\ket^{\rm c}\n
	=&
	-\ri \int \dd y \Bigg(
	\int_{-\infty}^{-y} \dd x\,\bra [q_{i,\alpha}(y), q_{j,\alpha}(x+y)]\Or\ket^{\rm c}
	-\int_{y}^\infty\dd x\, \bra[q_{i,\alpha}(y-x), q_{j,\alpha}(y)]\Or\ket^{\rm c}\Bigg)\n
	=&
	-\ri \int \dd x\, \Bigg(
	\int_{-\infty}^{-x} \dd y\,\bra [q_{i,\alpha}(y), q_{j,\alpha}(x+y)]\Or\ket^{\rm c}
	-\int_{-\infty}^x \dd y\,\bra [q_{i,\alpha}(y-x), q_{j,\alpha}(y)]\Or\ket^{\rm c}\Bigg)\n
	=&
	-\ri \int \dd x\,  \bra [q_{i,\alpha}(0), q_{j,\alpha}(x)] A(x)\ket^{\rm c}
	\label{result1}
\end{align}
where in the last line we use homogeneity of the state, and we have defined
\beq
	A (x) = \int_{-\infty}^{-x} \dd y\,\Or(-y\v e_\alpha) - \int_{-\infty}^x \dd y\,\Or((x-y)\v e_\alpha).\label{stepA}
\eeq
This quantity, and the steps leading to \eqref{setpAl} below, make sense within $\int \dd x\,\bra[q_i(0),q_j(x)] \cdot\ket^{\rm c}$, as $\int \dd x\,[q_i(0),q_j(x)]$ is a local observable supported around 0, and thanks to clustering. 

As in previous sections, $\Or(\v z)$ is the translation of $\Or$ by the vector $\v z$, and $\v e_\alpha$ is the unit vector in the $\alpha$ direction. Now let us consider the longitudinally integrated observable $O_\|=\int \dd z\,\Or(z\v e_\alpha)$. By clustering, the quantity $\bra (j_{ij,\alpha}+j_{ji,\alpha})O_\|\ket^{\rm c}$ is finite, and by the calculation above, it equates to
\beq
	\bra (j_{ij,\alpha}+j_{ji,\alpha})O_\|\ket^{\rm c}
	= -\ri\lim_{L\to\infty}  \int \dd x\,  \bra [q_{i,\alpha}(0), q_{j,\alpha}(x)] A_L(x)\ket^{\rm c}
\eeq
where
\beqa \label{U_L}
	A_L(x) &=& \int_{-L}^L \dd z\, \Bigg(\int_{-\infty}^{-x} \dd y\,\Or((z-y)\v e_\alpha) - \int_{-\infty}^x \dd y\,\Or((z+x-y)\v e_\alpha)\Bigg) \n
	&=& \int_{-L}^L \dd z\, \Bigg(\int_{-\infty}^{-x-z} \dd y\,\Or(-y\v e_\alpha) - \int_{-\infty}^{-z} \dd y\,\Or(-y\v e_\alpha)\Bigg) \n
	&=& -\int_{-L}^L \dd z\, \int_{z}^{x+z} \dd y\,\Or(y\v e_\alpha) \n
	&=& -\int_{-L}^{x+L}\dd y \int_{y-x}^y \dd z\, \Or(y\v e_\alpha) \n
	&=& -x\int_{-L}^{x+L}\dd y \, \Or(y\v e_\alpha). \label{setpAl}
\eeqa
In \eqref{result1} the integral over $x$ is supported on a finite region around 0, due to the locality of the conserved densities in the observable $[q_{i,\alpha}(0), q_{j,\alpha}(x)]$. Locality of this commutator is immediate in quantum systems, but perhaps not as well known in classical systems; we discuss a classical statement in appendix \ref{classical_appendix}. Understood within the correlation function in \eqref{result1}, we can take the limit $L\to\infty$ of \eqref{U_L}, which exists by clustering and gives
\beq
	\lim_{L\to\infty} A_L(x) = -x O_\|.
\eeq
Combining the expressions we have
\beq
	\bra (j_{ij,\alpha}(0)+j_{ji,\alpha}(0))O_\|\ket^{\rm c} = 
	\ri\int \dd^d x\, x^\alpha \bra [q_{i,\alpha}(0), q_{j}(\v{x})] O_\|\ket^{\rm c}.
\eeq
Referring to the definition of the marginal currents \eqref{transverse}, and likewise for the charge densities, we see that the above is equivalent to
\beq
	\bra (j_{ij}^\alpha(0)+j_{ji}^\alpha(0))O\ket^{\rm c} = 
	\ri\int \dd^d x\, x^\alpha \bra [q_{i}(0), q_{j}(\v{x})] O\ket^{\rm c}
\eeq
where $O$ is the total spatial integral of $\Or$,
\beq
	O = \int \dd^d x\,\Or(\v x).
\eeq
This expression holds for any operator $O$ which is an integrated local density. In particular, due to the tangent-manifold relation \eqref{tangent} we can take $O=Q_k$, leading to
\beq\label{stepdbeta}
	\frc{\p}{\p\beta^k} \Bigg(
	\bra \v{j}_{ij}+\v{j}_{ji}\ket
	-\ri\int \dd^d x\, \v{x}\bra [q_{i}(0), q_{j}(\v{x})]\ket^{\rm c}\Bigg) = 0.
\eeq
Thanks to the choice of gauge \eqref{gauge}, and as $\bra [q_i(0), q_{j}(\v{x})]\ket_{(\beta^\bullet=0)}=0$ in the infinite temperature state (a trace state), we obtain the desired result \eqref{firstmoment}.

\section{Proof of EKMS under the hypotheses of section \ref{sectThermo}}\label{sectproof}

The KMS condition \eqref{kms} for the state imposes conditions on the current observables of the theory. In this section we use \eqref{firstmoment} and the KMS condition to show the EKMS relation for the free energy fluxes, under the hypotheses of subsection \ref{ssectMES}. First, recall that for each conserved quantity $Q_k$ we have an associated free energy flux vector $\v g_k$, which by \eqref{free_flux} generates the currents $\v j_{ki}$ by differentiation with respect to the potentials $\beta^i$ defined by \eqref{tangent}. We first use the involution of the charges to associate a ``generalised time" $t^k$ parameterising the flow generated by each charge $Q_k$. We will denote the vector of potentials $\beta^{1_\alpha}$ as the vector $\v \beta^1$. Recall that these are associated to the conserved charges for space translation (if any) in the directions $x^\alpha$, the momentum operators $P_\alpha = Q_{1_\alpha}$. For models where there is no conserved momentum (such as lattice models), one simply sets $\v\beta^1=\v 0$ below.  Note that by convention, $t^{1_\alpha} = -x^\alpha$. Below the vector symbol, such as in $\vec \beta$, represents the set of all index values except $\{1_\alpha\}$.

The generalised times allow us to write the KMS condition, say in the form \eqref{kms2}, for a pair of a local conserved density $q_i$ and local observables $\Or$ as
\beq\label{qoq}
	\bra[q_i(\v{x},\vec t\,),\Or(0,0)]\ket = \bra q_i(\v{x},\vec t\,)\Or(0,0)\ket - \bra q_i(\v{x}+\ri \v \beta^1,\vec t-\ri \vec \beta\,)\Or(0,0)\ket
\eeq
Re-introducing explicit summations for clarity, this is evaluated as:
\beqa
	\bra[q_i(\v{x},\vec t\,),\Or(0,0)]\ket
	&=& -\int_0^1 \dd s\,\p_s \bra q_i(\v{x}+\ri \v \beta^1 s,\vec t - \ri \vec \beta s)\Or(0,0)\ket\n
	&=& \int_0^1 \dd s\,\sum_k {\ri \beta^k}\frc{\p}{\p t^k} \bra q_i(\v{x}+\ri \v \beta^1 s,\vec t - \ri  \vec \beta s)\Or(0,0)\ket\n
	&=& -\int_0^1 \dd s\,\sum_k {\beta^k} \bra [Q_k,q_i(\v{x}+\ri \v \beta^1 s,\vec t - \ri \vec \beta s)]\Or(0,0)\ket\n
	&=& -\int_0^1 \dd s\,\sum_k {\ri \beta^k}\nabla \cdot \bra \v{j}_{ki}(\v{x}+\ri \v \beta^1 s,\vec t - \ri  \vec \beta s)\Or(0,0)\ket.
\eeqa
In the classical case, using \eqref{poisson}, we obtain instead
\beq
	\bra \{q_i(\v{x},\vec t\,),\Or(0,0)\}\ket = \sum_k {\beta^k}\nabla \cdot \bra \v{j}_{ki}(\v{x},\vec t\,)\Or(0,0)\ket \quad\mbox{(classical case)}.
\eeq
We now choose $\Or = q_j$ and $\vec t=\vec 0$, and multiply by $\v{x}$ and integrate $\v{x}$ over $\R^d$; the integral exists by clustering of the state, and in particular the assumption that \eqref{clus} holds uniformly for $\tau_{-\ri  s}\Or_1(\v x)$ with $s\in[0,1]$. By invariance of the state, we can replace $\bra\cdots\ket$ by the connected correlators $\bra\cdots\ket^{\rm c}$ in \eqref{qoq}, and this gives
\beqa
	\int \dd^d x\,\v{x}\bra[q_i(\v{x}),q_j(0)]\ket &=& \int_0^1 \dd s\sum_k\ri \beta^k
	\bra \v{j}_{ki}(\ri \v \beta^1 s, - \ri \vec \beta s) Q_j\ket^{\rm c} \n
	&=& \sum_k\ri \beta^k
	\bra \v{j}_{ki}(\v 0,\vec 0\,) Q_j\ket^{\rm c} \n
	&=& -\sum_k\ri \beta^k
	\frc{\p}{\p \beta^j} \bra \v{j}_{ki}\ket 
\eeqa
where in the second equality we have used the stationarity of the state under the $Q_i$ and involution of the charges. Now as the left hand side is given by the identity \eqref{firstmoment} we can combine the expressions to get (this holds both in the quantum and classical cases)
\beq
	\bra \v{j}_{ij} + \v{j}_{ji}\ket = -\sum_k\beta^k \frc{\p}{\p\beta^j}\bra \v{j}_{ki}\ket
	= -\sum_k\frc{\p}{\p\beta^j} \big(\beta^k\bra \v{j}_{ki}\ket\big) + \bra \v{j}_{ji}\ket.
\eeq
We re-express the currents in terms of the free energy fluxes $\v{g}_i$, whereupon the expression reads
\beq
	\frc{\p \v{g}_i}{\p\beta^j} = -\sum_k\frc{\p}{\p\beta^j} \left(\beta^k\frc{\p \v{g}_k}{\p\beta_i}\right)
\eeq
which implies
\beq
	\sum_k\beta^k\v{g}_k = \sum_k\beta^k\v{F}_k+\v{G}
\eeq
where $\v{F}_k$ and $\v{G}$ are independent of the potentials. The $\v{F}_k$ may be set to 0, as the free energy fluxes are defined only up to a constant. This shows the EKMS relation \eqref{gkms1}.

%

\section{Conclusion}

In this paper we have obtained a relation between the free energy fluxes that holds in short-range many-body models of arbitrary dimension, under very general hypotheses. This gives general equations constraining average currents in Gibbs-like states, including generalised Gibbs ensembles, and clarifies properties of the entropy current. It can be seen as a constraint on the form of Euler-scale hydrodynamic equations, if they arise as emergent dynamical equations from an underlying many-body description.

The main result is established using the Kubo-Martin-Schwinger (KMS) relation. The KMS relation is a fundamental characteristic of (generalised) Gibbs states. We derived various implications. In particular we showed how it guarantees that the general Euler-scale hydrodynamic equations in space-varying fields, written in a model-independent way, have physically sound properties: their stationary state has the correct local-density approximation form, and they conserve entropy. This can be seen as an extension of the results of \cite{toth2003onsager,grisi2011current} which showed (in one dimension) that the Euler-scale Onsager relations (the symmetry of the $B$ matrix) guarantee physically sound properties of Euler-scale hydrodynamic equations in constant fields.

In general, many fundamental constraints may exist on the dynamics of emergent degrees of freedom, such as those found here, which encode the fact that they arise from an underlying short-range many-body system. It is important to establish the full set of such constraints. For instance, those on the Onsager matrix characterising diffusive hydrodynamics should help understand thermalisation in space-varying external fields. The present work also makes clear the importance of considering all conserved quantities at once, and in particular the flows they generate. This is particularly relevant in integrable systems, as these admit an extensive amount of conserved quantities, but it is also important in generic hydrodynamic equations, independent of any underlying integrable structures. It will be interesting to obtain the general theory where a fully symmetric treatment of all conserved quantities is recovered, and its physical consequences on physics away from equilibrium. Finally, it would be interesting to extend this to non-commuting conserved flows.

\medskip

\noindent
{\bf \em Acknowledgements}

\noindent We thank Gunter Sch\"{u}tz and Herbert Spohn for comments on the manuscript, in particular we are grateful to Gunter Sch\"{u}tz for pointing out the reference \cite{toth2003onsager}. BD also thanks Takato Yoshimura for discussions and for sharing related ideas. JD acknowledges funding from the EPSRC Centre for Doctoral Training in Cross-Disciplinary Approaches to Non-Equilibrium Systems (CANES) under grant EP/L015854/1.

\appendix

\section{Rigorous results in quantum spin chains} \label{app_rigour}

Our results are based on the properties \eqref{homo}, \eqref{clus}, \eqref{kms} and \eqref{tangent}, as well as the other assumptions made in subsection \ref{ssectMES}. These properties and assumptions can be  established in various families of models, for which the main results of section \ref{sectMain} are in fact mathematically rigorous. In order to illustrate this, here we explain how full mathematical rigour is indeed obtained in translation-invariant quantum spin chains with (finite local space and) finite-range interaction; this is the case $d=1$ and $x\in\Z$ of the discussion in the main text.

The most powerful rigorous setup for quantum spin chains, and more generally quantum lattice models, is that of uniformly hyperfinite (UHF) $C^*$-algebras, see \cite{IsraelConvexity,BratelliRobinson12}. See in particular \cite[Chap 6.2]{BratelliRobinson12}.

\subsection{Main framework}\label{ssectrigmain}

The quantum chain is characterised by a UHF algebra $\mathfrak U$, which may be seen as the completion with respect to the operator norm of the algebra $\mathfrak L = ({\rm End}\,\C^2)^{\Z}$ of local observables on $\Z$. Space translations form a representation of $\Z$ on algebra $*$-automorphisms. We assume that the model possesses a certain number of extensive, homogeneous (translation-invariant) conserved charges in involution. It is sufficient to provide the associated conserved densities and currents, so we are given $q_i(x),j_{ki}(x)\in \mathfrak L$ for $k,i$ in some finite index set $I$ and $x\in\Z$, such that relations \eqref{evol} are satisfied, and that $q_i(x), j_{ki}(x)$ are $x$-translates of $q_i(0),j_{ki}(0)$. In particular, in \eqref{evol} the commutator is a local observable, $[Q_k, q_i(x)]\in\mathfrak L$, as only a finite number of terms in the extensive charge \eqref{Qi} contribute.

For concreteness, one may take the Heisenberg spin chain, whose Hamiltonian density is expressed in terms of the Pauli matrices $\vec\sigma_x$ acting on sites $x\in\Z$,
\beq
	q_2(x) = \vec\sigma_x\cdot\vec\sigma_{x+1}.
\eeq
This model possesses a large number of extensive conserved charges thanks to its integrability. In particular, the total spin in every direction is conserved. We may choose one direction, and set the density with label 0 to
\beq
	q_{0}(x) = \sigma_x^{\rm z}.
\eeq
Another conserved charge is the one with density
\beq
	q_4(x) = \vec\sigma_x\cdot (\vec\sigma_{x+1}\times\vec\sigma_{x+2}).
\eeq
It turns out that this is (proportional to) the energy current $j_2(x)$, although this will not play any role in our discussion. By transfer matrix methods, it is simple to find more and more local conserved densities, but for concreteness it is sufficient to concentrate on these, with $I=\{0,2,4\}$.
Direct calculations also give the explicit forms of $j_{ki}$ for $k,i\in\{0,2,4\}$.

As we do not have a momentum operator in quantum spin chains, equations \eqref{stress} and \eqref{j1i} do not make sense and are omitted; this does not affect the main results.

We consider a set of parameters $\beta^i\in\R$ for $i\in I$ and construct the formal extensive charge $W=\sum_{i\in I} \beta^i Q_i$ as in \eqref{WQi}, where the sum over $i$ is finite. Again, the generator $\ri [W,\cdot]$ is well defined on $\mathfrak L$, with $\ri [W,\cdot]:\mathfrak L\to\mathfrak L$. As $\mathfrak L$ is dense in $\mathfrak U$, the closure of this generator generates a unique one-parameter strongly continuous group of $*$-automorphisms $\tau:t\in\R\to\tau_t$. Here finiteness of the interaction range, that is the fact that $\ri [W,\cdot]:\mathfrak L\to\mathfrak L$, is used, although weaker conditions are possible. Then, we consider a set of states $\bra\cdots\ket$, positive linear functionals $\mathfrak U\to\C$, parametrised (implicitly) by $\beta^i\in\R$ for $i\in I$. These are defined by the KMS relation \eqref{kms} for every analytic element $\Or\in\mathfrak U$ (entire analytic elements of $\mathfrak U$ form a dense subspace). The state is unique for every choice of finite $\beta^i$, and it is the trace state if $\beta^i=0 \;\forall\;i$. See \cite[Chap 6.2]{BratelliRobinson12}. In fact, in a KMS state, the two-point function $\bra \tau_t \Or_2 \Or_1\ket $ is analytic in $t\in[0,-\ri)$ for any $\Or_1,\Or_2\in\mathfrak U$, and its boundary value satisfies the KMS relation \eqref{kms}  \cite{Arakitime}.

Further, by \cite{Araki}, the state is invariant under space translations, Eq~\eqref{homo}, and hence, thanks to \eqref{evol}, it is stationary under every generator $[Q_i,\cdot]$. Also, as a result of \cite{Araki} again, the state is exponentially clustering, Eq.~\eqref{clus}. More precisely, one may use the statement from \cite[Thm 6.1]{Doyon2017}, restated as follows: Let $K\subset \R^{|I|}$ be a compact subset and $p>1$. There exists $\nu,a>0$ such that, for every multiplet $\beta^\bullet\in K$ and every $\Or,\Or'\in\mathfrak L$,
\beq\label{clusrig}
	|\bra \Or(x)\Or'(0)\ket^{\rm c}|\leq
	\nu |\Or|^a |\Or'|^a \big({\rm dist}(\Or(x),\Or'(0))\big)^{-p}
\eeq
where $|\Or|$ is the size of the support of $\Or(x)$ (it is independent of $x$), and ${\rm dist}(\Or(x),\Or(0))$ is the distance between the supports of $\Or(x)$ and $\Or'(0)$ (it grows like $|x|$ as $x\to\pm\infty$); see \cite{Doyon2017} for the details. This is the statement of algebraic clustering for every power $p$, instead of exponential clustering, but this is sufficient.

As parts of the assumptions of subsection \ref{ssectMES}, clustering is required not only for local observables, but also for imaginary-time evolved ones, with $\Or(x)$ replaced by $\tau_{-\ri s} \Or(x)$ for $s\in[0,1]$. We are not currently aware of a study of clustering of such imaginary-time evolved local observables in thermodynamic states, except for \cite{DoyonProjection}. There, it is shown that, for complex time $t$ with $|t|$ small enough, $t$-evolved local observables are analytic, and clustering for $\bra\tau_t\Or(x)\Or'(0)\ket^{\rm c}$ holds uniformly. Thus, \eqref{clusrig} holds with $\Or(x)$ replaced by $\tau_{-\ri s} \Or(x)$ for $s\in[0,1]$, for all $\beta^\bullet$ in some neighbourhood of 0. The neighbourhood of 0 depends on the particular model under study. This will therefore establish the EKMS relation for $\beta^\bullet$ in a neighbourhood of 0. But by analyticity of the state in $\beta^\bullet$ (see below), the relation stays valid for all $\beta^\bullet\in\R^{|I|}$; hence this is sufficient.

We believe that it is possible to establish directly clustering of $\bra\tau_{-\ri s} \Or(x)\Or'(0)\ket^{\rm c}$ for all $\beta^\bullet$ by using the results of \cite{KGKRE14}; however this would require a more careful analysis of the infinite-volume limit.

Finally, \cite[Thm 6.1]{Doyon2017} establishes part of the tangent-manifold relation \eqref{tangent}, and can be extended to the full relation\footnote{Recall that for discrete space, the symbol $\int \dd x$ is meant as a discrete sum $\sum_{x\in\Z}$.}. The theorem is concerned with the case with $\beta^i=0$ for all $i\neq 2$, that is, the case  $W = \sum_{i\in I}\beta^i Q_i = \beta^2 Q_2\equiv \beta Q_2$; it shows in this case that the derivative $\dd \bra\Or\ket/\dd\beta = -\bra \Or Q_2 \ket^{\rm c}$ exists and is analytic in $\beta$, for every $\Or\in\mathfrak L$. It is a simple matter to extend the proof to any finite set $I$, by generalising the steps \cite[Eq 70-77]{Doyon2017} as follows.

The proof is based on expressing the KMS state $\bra\cdots\ket$ as an infinite volume limit, and on \cite[Thm 2]{KGKRE14}. In \cite[Eq 66]{Doyon2017}, the finite volume version of the system, on which the limit is taken, was defined with ``open boundaries". First, we need to reformulate this to ``periodic boundary conditions", in order to ensure that $Q_i$, in the finite volume version, are in involution. The choice of boundary condition does not affect this limit, as shown for instance by \cite[Cor 2]{KGKRE14}, and in particular all conditions of \cite[Thm 2]{KGKRE14} still hold.

Let $\mathfrak L_{[-N,N]} = ({\rm End}\,\C^2)^{[-N,N]\cap \Z}$ be the space of local observables supported on the finite set of sites $[-N,N]\cap \Z$. For all $N$ large enough, there is $n>0$ such that $q_i(x), j_{ki}(x)\in\mathfrak L_{[-N,N]}$ for all $x\in[-N+n,N-n]$ and all $i,k\in I$. Let us define new observables, still denoted $q_i(x), j_{ki}(x)$, that belong to $\mathfrak L_{[-N,N]}$ for $x\in[-N,N]$, as translates of $q_i(0),j_{ki}(0)$, under translation automorphisms that are periodic on $[-N,N]\cap\Z$. The new $q_i(x), j_{ki}(x)$ may differ from the old ones only ``near the boundaries", for $x\in[-N,-N+n-1]\cup [N-n+1,N]$. Then, the limit
\beq
	\bra\Or\ket = \lim_{N\to\infty} \frc{
	\Tr_{\mathfrak L_{[-N,N]}}\Big(
	e^{-\sum_{i\in I}\beta^iQ_i^{(N)}} \Or\Big)
	}{
	\Tr_{\mathfrak L_{[-N,N]}}\Big(
	e^{-\sum_{i\in I}\beta^iQ_i^{(N)}} \Big)
	},\quad Q_i^{(N)} = \sum_{x\in[-N,N]} q_i(x)
\eeq
exists for every $\Or\in\mathfrak L$ and gives, after completion to $\mathfrak U$, the KMS state $\bra\cdots\ket$ with the properties discussed above. 

As, by construction, \eqref{evol} holds for $Q_i$ replaced by $Q_i^{(N)}$, we have $[Q_i^{(N)},Q_j^{(N)}]=0$. Then, the steps \cite[Eq 70-77]{Doyon2017} follow, with the derivative $\dd/\dd\beta$ replaced by $\p/\p\beta^i$ for any $i\in I$, and the trace state replaced by the state where $\beta^i=0$. The arguments presented there show that $\p\bra \Or\ket/\p\beta^i$ is analytic in $\beta^j\in\R$ for every $j\in I$ and  fixed $\beta^{k\neq j}$ (the argument is that the finite-volume version is analytic, and analyticity subsists in the infinite-volume limit thanks to the uniform clustering of \cite[Thm 2]{KGKRE14}); hence, in particular, it is continuous. Therefore,  \eqref{tangent} holds, and the derivative is continuous.

Thus, all assumptions of subsection \ref{ssectMES}, at the basis of the general results, are satisfied in translation-invariant quantum spin chains with finite-range interactions. The Heisenberg chain, with the choice $I=\{0,2,4\}$ as above, gives an explicit and nontrivial example.

\subsection{Free energy fluxes}

As the main results are concerned with the free energy fluxes $g_k$, it is important to guarantee that these exist and satisfy the required relations. As explained in subsection \ref{ssectfluxes}, this is a result of standard arguments, and we simply repeat these arguments here in the context of quantum spin chains for completeness and full mathematical rigour.

First, the symmetry $\mathsf B_{kij} = \mathsf B_{kji}$ follows from translation invariance and clustering, and it is easy to check that the clustering statement \eqref{clusrig} is sufficient in order for the derivation in appendix \ref{app_symmetry_B_mat} to hold rigorously. Then, the equality
\beq
	\p\bra j_{ki}\ket/\p\beta^j = \p\bra j_{kj}\ket/\p\beta^i,
\eeq
which follows from this symmetry, along with continuity of the derivatives, established above, show the existence of a second-differentiable function $g_k$ such that \eqref{free_flux} holds. As $\bra j_{ki}\ket$ are analytic in $\beta^j$ for every $j$, so are $g_k$. Similar arguments hold for the existence of a free energy $f$, using symmetry of the $\mathsf C_{ij}$ matrix. Hence, the discussion in subsection \ref{ssectfluxes} is mathematically rigorous (again, omitting \eqref{g1} as we do not assume the presence of a conserved momentum in quantum spin chains).

Assuming that the main result \eqref{gkms1} holds, it is also a simple matter to verify that Eqs.~\eqref{gkmsent}-\eqref{betabetaj}, except \eqref{fT}, are therefore rigorously established.

\subsection{Proof of the EKMS relation and a theorem}

In section \ref{sectmoment} we establish the general identity \eqref{firstmoment}, which plays an important role in the derivation of the main result, the EKMS relation \eqref{gkms1}. It is a easy to see that all steps are fully rigorous. In one dimension, it is not necessary to define the transverse marginals \eqref{transverse}, simplifying the steps slightly. In particular, in \eqref{step11} and \eqref{result1} the series are convergent thanks to locality of the conserved densities and clustering. We note for instance that the quantity $\int \dd x\,[q_i(0),q_j(x)]$ is local: supported on a finite number of sites around 0. Equation \eqref{stepA} and the steps leading to \eqref{setpAl} are rigorous within $\int \dd x\,\bra[q_i(0),q_j(x)] \cdot\ket^{\rm c}$. Finally locality of $\int \dd x\,[q_i(0),q_j(x)]$ allows us to use the tangent-manifold relation and obtain \eqref{stepdbeta}.

In section \ref{sectproof} the main result is obtained. Clustering \eqref{clusrig}, in particular analyticity and clustering of imaginary-time evolved observables for $\beta^\bullet$ near enough to 0, is established rigorously; so are the first-moment relation \eqref{firstmoment}, the tangent-manifold relation \eqref{tangent}, and the trace property of the infinite-temperature state. This guarantees that every step is rigorous.

Thus, we have established \eqref{gkms1}, and hence Eqs.~\eqref{gkmsent}-\eqref{betabetaj} (except \eqref{fT}), in translation-invariant quantum spin chains with finite-range interactions, when a finite number of extensive conserved charges $Q_i,\,i\in I$ with local (i.e.~finite-range) densities are considered. The arguments presented hold for $\beta^\bullet$ near enough to 0, where clustering at large distances of imaginary-time evolved observables has been proven in the literature, but by analytic continuation all relations still hold for $\beta^\bullet\in\R^{|I|}$. Again, the Heisenberg chain, with the choice $I=\{0,2,4\}$ as in subsection \ref{ssectrigmain}, gives an explicit and nontrivial example. It is simple to show that the Heisenberg chain has parity symmetry, hence in this case $G=0$.

We gather the main result in the following theorem.

\begin{theorem}\label{theochain} Let $\bra\cdots\ket$ be $(\tau,1)-KMS$ state in a quantum spin chain on infinite volume, parametrised by $\beta^i\in\R$ for $i\in I$, where $I$ is a finite set. Take $\tau$ to be generated by $W = \sum_{i\in I} \beta^i Q_i$, where $Q_i$ are extensive conserved quantities in involution, with local (supported on finite numbers of sites) densities $q_i(x)$ and associated traceless local generalised currents $j_{ki}(x)$. Then there are differentiable real functions $g_k:\R^{|I|}\to\R$ of $\beta^i$'s such that $\bra j_{ki}\ket = \p g_k/\p\beta^i$, and there exists $G\in\R$ such that \eqref{gkms1} holds, $\sum_{k\in I} \beta^kg_k = G$ for all $\beta^\bullet \in \R^{|I|}$. In particular, relations \eqref{gibeta}, \eqref{swap_indices} and \eqref{betabetaj} hold for all $\beta^\bullet \in \R^{|I|}$.
\end{theorem}


\section{KMS and tangent-space relations}\label{appkmstangent}

The KMS and tangent-space relations \eqref{kms} and \eqref{tangent}, respectively, are both important characteristics of (generalised) Gibbs states. The KMS relation can be taken as a fundamental definition of such states, see \cite{IsraelConvexity,BratelliRobinson12,aizenman1977equivalence}. The tangent-space relation is perhaps less universal, as it is expected to hold only in regions of the manifold of states where correlations decay fast enough. However, at large enough temperatures, it can be proven rigorously \cite{Doyon2017,KGKRE14}, where the conserved quantity $Q_i$ in \eqref{tangent} is made rigorous via the concept of (linearly) extensive charges. In this appendix, we briefly explain how both relations are compatible, under the relation \eqref{WQi}.

The assumption that the conserved charges $Q_i$ are in involution gives, from \eqref{WQi} and \eqref{tau_s}, the relation (setting $\hbar=1$)
\beq
	-\frc{\p}{\p\beta^i} (\tau_s \Or) = -\ri s [Q_i,\tau_s \Or].
\eeq
Clearly, the state depends on all $\beta^i$'s. In order to clarify the calculation, let us temporarily use the following notation
\beq
	\bra\cdots\ket = \omega(\cdots).
\eeq
Let us assume that differentiating the state with respect to $\beta^i$ inserts an extensive charge. Let us denote this charge by $\t Q_i$:
\beq
	-\Big(\frc{\p}{\p\beta^i}\omega\Big)(\cdots) = \omega(\t Q_i\cdots ) - \omega(\t Q_i )\omega(  \cdots)
	= \bra \t Q_i \cdots\ket^{\rm c}
\eeq
where the superscript c means that the connected correlation function is taken with respect to the extensive charge only. Various results point to the fact that a state derivative must be representable as the insertion of an extensive charge, see \cite{Doyon2017}; although we do not know of a complete proof.  The charge does not in fact need to be conserved, but for simplicity we will assume that the state $\omega$ is invariant under $\t Q_i$. Under these assumptions, reverting to the bracket notation and differentiating the KMS relation \eqref{kms}, we obtain
\beq
	\bra \t Q_i\Or_1 \Or_2 \ket^{\rm c} = 
	\bra \t Q_i\tau_{-\ri} \Or_2 \Or_1 \ket^{\rm c}
	- \bra[Q_i,\tau_{-\ri} \Or_2] \Or_1\ket.
\eeq
We can write
\beq
	\bra[Q_i,\tau_{-\ri} \Or_2] \Or_1\ket = \bra Q_i\tau_{-\ri} \Or_2 \Or_1\ket^{\rm c}
	- \bra\tau_{-\ri} \Or_2Q_i\Or_1\ket^{\rm c}
\eeq
and using the KMS relation again as well as invariance of the state under the actions of $Q_i$ and $\t Q_i$,  we obtain
\beq
	\bra [(\t Q_i - Q_i),\Or_1] \Or_2\ket^{\rm c} = 0.
\eeq
Since this holds for all $\Or_2$, it implies, under general conditions, that $[(\t Q_i - Q_i),\Or_1]=0$. Since this holds for all $\Or_1$, the difference $\t Q_i - Q_i$ must lie in the centre of the operator algebra. Again under general conditions, this means $\t Q_i - Q_i\propto \bf 1$. Since the identity operator $\bf 1$ trivially clusters, then $\t Q_i$ is defined only up to additions by multiples of $\bf 1$, and thus we may take
\beq
	\t Q_i = Q_i.
\eeq

\section{Symmetry of the $B$-matrices}\label{app_symmetry_B_mat}
The symmetry of the $\v B$ matrix in one dimension is well established \cite{toth2003onsager,grisi2011current,PhysRevX.6.041065,dNBD2,KarevskiCharge2019}. The higher dimensional context is also discussed in \cite{KarevskiCharge2019}. In this section we give a short derivation of the symmetry of the $\v B$-matrix  in general dimension, within the context as described in section \ref{sectThermo}. The derivation follows from standard dimensional reduction techniques, as suggested in \cite{KarevskiCharge2019}.

We first use invariance of the state under spatial translation and the action of the charges $Q_i$ to write
\beqa
    \nabla\cdot\bra q_{i}(\v{x},\{t^\ell\})\v{j}_{jk}(0,\{0\})\ket^{\rm c}&=&\nabla\cdot\bra q_{i}(0,\{0\})\v{j}_{jk}(-\v{x},\{-t^\ell\})\ket^{\rm c}\nonumber \\
    &=&-\p_{t^j}\bra q_{i}(0,\{0\})q_k(-\v{x},\{-t^\ell\})\ket^{\rm c}\nonumber \\
    &=&-\p_{t^j}\bra q_{i}(\v{x},\{t^\ell\})q_k(0,\{0\})\ket^{\rm c}\nonumber \\
    &=&\nabla\cdot\bra \v{j}_{ji}(\v{x},\{t^\ell\})q_k(0,\{0\})\ket^{\rm c}. \label{eqapp}
\eeqa
Let us integrate both sides along the hyperplane spanned by $x^2,\ldots,x^d\in\R^{d-1}$. By clustering, the integration along the asymptotic boundary of this hyperplane vanishes, and there remains
\beq
	\frc{d}{dx^1} \int \prod_{\alpha=2}^d \dd x^\alpha\, \big(\bra q_{i}(\v{x})\v{j}_{jk}(0)\ket^{\rm c} - 
	\bra \v{j}_{ji}(\v{x})q_k(0)\ket^{\rm c}\big) = 0
\eeq
where we set all times to $t^\ell=0$. Again by clustering, the integral vanishes in the limits $x^1\to\pm\infty$. As the derivative with respect to $x^1$ is zero throughout, the integral must vanish identically. That is,
\beq
	\int \prod_{\alpha=2}^d \dd x^\alpha\, \big(\bra q_{i}(\v{x})\v{j}_{jk}(0)\ket^{\rm c} - 
	\bra \v{j}_{ji}(\v{x})q_k(0)\ket^{\rm c}\big) = 0.
\eeq
By clustering, the integral exists on both terms in the integrand separately, and further, the integral over $x^1\in\R$ of each resulting term exists. Thus we find
\beq
	\int \dd^{d}x\, \bra q_{i}(\v{x})\v{j}_{jk}(0)\ket^{\rm c} =
	\int \dd^d x \bra \v{j}_{ji}(\v{x})q_k(0)\ket^{\rm c}.
\eeq
Each side may be identified with an element of the $\v{B}_{j\bullet\bullet}$ matrix (using homogeneity of the state and, in the quantum case, invariance of the state under the flow generated by $Q_i$), giving its sought symmetry,
\begin{equation} \label{some_eq0}
    \v{B}_{jki}=\v{B}_{jik}.
\end{equation}
\section{Poisson brackets of charge densities}\label{classical_appendix}
In classical Hamiltonian particle systems it it possible to write explicit expressions for quasi-local observables such as conserved densities, and thereby to show that the expected decay of the Poisson bracket at large space separations occurs quite generically. For simplicity we present the results in one dimension, the generalisation to higher dimensions being straightforward. For the notion of locality of classical densities and currents, see also \cite{LLL77,SpoToda,DToda}.

The most general observable satisfying spatial-translation invariance can be written in terms of the canonical variables $\{x_n,p_n\}$ as 
\begin{equation}
    q_i(x)=\sum_n\delta(x-x_n)\sum_{N=0}^\infty\sum_{m_1\ne n}\cdots\sum_{m_N\ne n}h_i^{n,m_1,\cdots,m_N}(p_n;\{(x_n-x_{m_1},p_{m_1}),\cdots,(x_n-x_{m_N},p_{m_N})\}),
\end{equation}
where we consider implicit expressions integrated against some test function over $x$. The notion of quasi-locality can be viewed as a condition of connectedness of the functions
\begin{equation}
    h_i^{n,m_1,\cdots,m_N}(p_n;\{(x_n-x_{m_1},p_{m_1}),\cdots,(x_n-x_{m_N},p_{m_N})\})\sim e^{-|x_n-x_{m_\ell}|^\alpha}\;\mathrm{for}\;\;|x_n-x_{m_\ell}|\gg 1,
\end{equation}
for some $\alpha>0$. Examples of systems with charge densities of this form are generic systems with only particle number, momentum and the Hamiltonian conserved, and the charge densities of the Toda gas obtained by taking derivatives of the Lax-matrix \cite{Fla74,SpoToda,DToda}.

To calculate the Poisson bracket we first write, using obvious condensation of notation, the results
\begin{align}
    \p_{x_\ell}q_i(x)=&\sum_n\sum_{N}\sum_{\vec{m}\ne n}\bigg[\delta(x-x_n)\p_{x_\ell}-\delta_{n\ell}\delta'(x-x_n)\bigg]h_i^{n,\vec{m}}(p_n;\{(x_n-x_{\vec{m}},p_{\vec{m}})\}),
\end{align}
and
\begin{equation}
    \p_{p_\ell}q_i(x)=\sum_{n}\delta(x-x_n)\sum_{N}\sum_{\vec{m}\ne n}\p_{p_\ell}h_i^{n,\vec{m}}(p_n;\{(x_n-x_{\vec{m}},p_{\vec{m}})\}).
\end{equation}
The Poisson bracket is then found to be
\begin{align} \label{poisson_expansion}
    \{q_i(x),q_j(0)\}=&\sum_\ell\sum_{nn'}\sum_{NN'}\sum_{\vec{m}\ne n}\sum_{\vec{m}'\ne n'}\nonumber \\
    \times&\bigg[\left[\left(\delta(x-x_n)\p_{x_\ell}-\delta_{n\ell}\delta'(x-x_n)\right)h^{n\vec{m}}_i(p_n,\{(x_n-x_{\vec{m}},p_{\vec{m}})\})\right]\nonumber \\&\;\times\delta(x_{n'})\p_{p_\ell}h^{n'\vec{m}'}_j(p_{n'};\{(-x_{\vec{m}'},p_{\vec{m}'})\})\nonumber \\&
    \;-\left[\left(\delta(-x_{n'})\p_{x_\ell}-\delta_{n'\ell}\delta'(-x_{n'})\right)h_j^{n'\vec{m}'}(p_{n'},\{(-x_{\vec{m}'},p_{\vec{m}'})\})\right]\nonumber \\&\;\times\delta(x-x_n)\p_{p_\ell}h^{n\vec{m}}_i(p_{n};\{(x_n-x_{\vec{m}},p_{\vec{m}})\})\bigg].
\end{align}
We now take $|x|\gg 1$, and consider the ball $\mathcal{B}_R(x)$ of radius $R<|x|/2$ but with $R\gg 1$, centered at position $x$. We consider likewise the ball $\mathcal{B}_R(0)$ centered at 0. We denote the set of particles contained within $\mathcal{B}_R(x)$ as $N_R(x)$, likewise for $\mathcal{B}_R(0)$, and as the balls are non-overlapping we truncate the sums over $n,N,\vec{m}$ and $n',N',\vec{m}'$ to contain only the particles in $N_R(x)$ and $N_R(0)$ respectively, obtaining errors of order $e^{-R^\alpha}$.

Now consider the two terms in the large square parentheses. In order for the derivative with respect to $p_\ell$ to be non-zero, particle $\ell$ must be contained within the sums over $n',N',\vec{m}'$ ($n,N,\vec{m}$) to contribute to the first (second) term. However, applying a similar argument to the multiplying terms, $l$ must be contained in the other sum $n,N,\vec{m}$ ($n',N',\vec{m}'$) to contribute. As the terms in each set of sums are disjoint, the contribution from at least one of the sums is 0, and we therefore have that the expression vanishes within the errors  of order $e^{-R^\alpha}$. In conclusion, the large $x$ behaviour of the commutator is of the order:
\begin{equation}
    \{q_i(x),q_j(0)\}\sim e^{-|x|^\alpha}\;,\;\;|x|\rightarrow \infty.
\end{equation}
Such an exponential decay is sufficient for the proof of section \ref{sectmoment}. For a purely local charge, the same derivation shows that the commutator vanishes outside some finite radius. This derivation readily applies to infinite systems, as long as they are at finite density.

\end{document}